\PassOptionsToPackage{table,dvipsnames}{xcolor}
\documentclass[sigconf]{acmart}

\setcopyright{none}
\renewcommand\footnotetextcopyrightpermission[1]{}
\usepackage{enumitem}
\usepackage{amsmath}
\usepackage{booktabs}
\usepackage{algorithm}
\usepackage{algorithmicx}
\usepackage{algpseudocodex}
\usepackage{hyperref}
\usepackage{colortbl}
\usepackage[capitalise]{cleveref}
\usepackage{url}
\usepackage{xspace}
\usepackage{xcolor}
\usepackage{graphicx}
\usepackage[caption=false,font=footnotesize]{subfig}
\usepackage{pifont}
\definecolor{darkgreen}{rgb}{0.0, 0.6, 0.0}
\newcommand{\cmark}{\color{darkgreen}\ding{51}}
\newcommand{\xmark}{\color{red}\ding{55}}
\usepackage{makecell}
\usepackage{arydshln}
\usepackage{tikz}
\tikzset{blobcell/.style={draw, minimum height=1.6em, inner xsep=6pt,
    outer sep=0pt, anchor=west, font=\small}}

\newcommand{\crefsectionfmt}[1]{%
  \crefformat{#1}{\S\,##2##1##3}%
  \Crefformat{#1}{\S\,##2##1##3}%
  \crefrangeformat{#1}{\S\S\,##3##1##4 to~##5##2##6}%
  \Crefrangeformat{#1}{\S\S\,##3##1##4 to~##5##2##6}%
  \crefmultiformat{#1}{\S\S\,##2##1##3}{ and~##2##1##3}{, ##2##1##3}{ and~##2##1##3}%
  \Crefmultiformat{#1}{\S\S\,##2##1##3}{ and~##2##1##3}{, ##2##1##3}{ and~##2##1##3}%
}
\crefsectionfmt{section}
\crefsectionfmt{subsection}
\crefsectionfmt{subsubsection}

\Crefname{appsec}{Appendix}{appendices}

\newcommand{\name}{\textrm{Apollo}\xspace}
\newcommand{\tname}{\textrm{T-Apollo}\xspace}
\newcommand{\hname}{\textrm{H-Apollo}\xspace}
\newcommand{\trustees}{trustees\xspace}
\newcommand{\nontrustees}{non-trustees\xspace}
\newcommand{\nontrustee}{non-trustee\xspace}
\newcommand{\trustee}{trustee\xspace}
\newcommand{\checksum}{tag\xspace}

\newcommand{\multiss}{\textrm{MLSS}\xspace}
\newcommand{\tmultiss}{\textrm{TMLSS}\xspace}
\newcommand{\hmultiss}{\textrm{HMLSS}\xspace}

\newcommand{\randomsample}{\stackrel{R}{\leftarrow}}

\newcommand{\fieldspace}{\mathbb{Z}_p}
\newcommand{\hash}[1]{\mathsf{H}({#1})}
\newcommand{\galois}[2]{GF({#1}^{#2})}

\usepackage{mathtools}

\DeclarePairedDelimiter\floor{\lfloor}{\rfloor}
\newcommand{\user}{\mathcal{U}}

\newcommand{\adversary}{\mathcal{A}}

\newcommand{\key}{\kappa}
\newcommand{\share}[1]{\key_{#1}}

\newcommand{\addrbook}{{A}}
\newcommand{\numaddrbook}{{n}}
\newcommand{\addrbookpart}[1]{a_{#1}}
\newcommand{\trusteesset}{{A}_T}
\newcommand{\numtrusteesset}{{n}_T}
\newcommand{\numtrusteessetonly}{\lfloor \delta{n}_T \rfloor}

\newcommand{\numtrusteessetrec}{{n}_{T,\mathrm{rec}}}
\newcommand{\numtrusteessetreconly}{{n}^{\gamma}_{T,\mathrm{rec}}}

\newcommand{\nullval}{\bot}

\newcommand{\packet}[1]{\mathrm{pkt}_{#1}}
\newcommand{\blob}[1]{\mathrm{blob}_{#1}}
\newcommand{\coord}[1]{({x}_i, {y}_i)}
\newcommand{\n}{{n}}

\newcommand{\sharesnum}{\nu}

\newcommand{\threshold}{\tau}
\newcommand{\numss}{\beta}
\newcommand{\spp}{\gamma}
\newcommand{\absolute}{\alpha}

\DeclareMathOperator{\obt}{obt}

\DeclareMathOperator{\totalnum}{total}
\DeclareMathOperator{\single}{single}
\DeclareMathOperator{\sub}{sub}

\newcommand{\salt}{\mathrm{salt}}
\newcommand{\Hofstuff}[2]{\mathsf{H}(#1\, \vert\vert\, #2)}

\DeclareMathOperator{\rec}{rec}
\DeclareMathOperator{\recovered}{recovered}
\newcommand{\ssmath}[1]{\mathrm{ss}_{#1}}
\DeclareMathOperator{\interp}{int}
\DeclareMathOperator{\relev}{relevant}

\begin{document}

\title{Towards Self-Recovery and Metadata Privacy in Social Vault Recovery
with \name}

\author{Shailesh Mishra}
\affiliation{%
  \institution{EPFL}
  \city{Lausanne}
  \country{Switzerland}}
\email{shailesh.mishra@epfl.ch}

\author{Simone Colombo}
\authornote{Work partially done while at EPFL.}
\affiliation{%
  \institution{King's College London}
  \city{London}
  \country{United Kingdom}}
\email{simone.colombo@kcl.ac.uk}

\author{Pasindu Tennage}
\authornotemark[1]
\affiliation{%
  \institution{Digital Asset}
  \city{Zurich}
  \country{Switzerland}}
\email{pasindu.tennage@gmail.com}

\author{Martin Burkhart}
\affiliation{%
  \institution{armasuisse}
  \city{Zurich}
  \country{Switzerland}}
\email{martin.burkhart@armasuisse.ch}

\author{Bryan Ford}
\affiliation{%
  \institution{EPFL}
  \city{Lausanne}
  \country{Switzerland}}
\email{bryan.ford@epfl.ch}

\renewcommand{\shortauthors}{Mishra et al.}

\begin{abstract}
Social recovery enables users to enlist trusted contacts,
or trustees,
to help recover lost access to end-to-end-encrypted repositories
or vaults.
However,
existing recovery mechanisms often make
strong memorability assumptions
about what users will remember.
Weakening these memorability assumptions
to increase the robustness of recovery
is possible,
but may leak sensitive metadata about the user and/or trustees
if done na\"ively.
This paper's first contribution is to draw attention to and formalize
this basic tension between \emph{memorability}
and \emph{metadata privacy} in social vault recovery.
Our second contribution is \emph{Apollo},
a social recovery mechanism intended to explore
how close we can jointly approach the ideals of
\emph{self-recovery} -- avoiding any memorability assumptions --
while strongly protecting recovery metadata privacy.
Apollo approximates the ideal of self-recovery
by relying only on a threshold of social \emph{reconnection} events,
which may be initiated either by the user or the user's contacts.
Apollo thereby has a chance of succeeding in vault recovery
even in a worst-case scenario
where the user has forgotten all metadata,
including even the vault's existence.
To protect the metadata's privacy,
Apollo distributes either real or fake (chaff) data
to all of a user's contacts, not just the user's trustees,
thus systematically anonymizing the trustees among the larger set of contacts.
To make this anonymity set scalable,
\name uses a novel \emph{multi-layered secret sharing} scheme
to mitigate the computational overhead of recovery in this setting,
which would otherwise be exponential in the recovery threshold.
Finally, we evaluate a prototype implementation of Apollo.
\name reduces the probability of malicious recovery to under $0.1\%$
for an adversary capable of obtaining shares from every 1-out-of-2 contacts.
After reconnecting with $30$ contacts,
the multi-layered design shows
an improvement of $5$ orders in computation time,
compared to a single-layered approach.
\end{abstract}



\makeatletter
\renewcommand*{\@fnsymbol}[1]{\textdagger}
\makeatother

\maketitle
\pagestyle{plain}
\thispagestyle{plain}

\section{Introduction}
\label{sec:intro}

Digital vaults,
such as password managers~\cite{1password,dashlane,bitwarden},
OS-managed storage services~\cite{icloud,onedrive},
and cryptocurrency wallets~\cite{coinbase,metamask,trezor},
play a key role in safeguarding digital assets.
To maximize privacy,
end-to-end-encrypted (e2ee) vaults%
~\cite{appleadvdataprot,keybase,hashicorp}
normally encrypt their content with a key
held solely by the user.
If a user loses her device(s) containing the key, however,
she may permanently lose access to her digital assets,
leading to dire consequences:
e.g., a man having to search
landfills for a lost Bitcoin key \cite{jameshowells},
or the IACR having to rerun an election after losing
a cryptographic key \cite{arstechnica2025iacr}.
Vault \emph{recovery} solutions are crucial
for guarding against
such irreversible loss
by backing up the encryption key.

Existing vault recovery solutions fall into two broad categories:
(i) \emph{custodial solutions}%
~\cite{dropbox,gdrive,icloud,onedrive,coinbase,redsafefaq},
where a trusted party would assist Alice with the recovery,
and (ii) \emph{self-sovereign solutions}%
~\cite{appleadvdataprot,keybase,hashicorp,trezor,ledger,safepal,metamask},
where Alice needs to handle the recovery on her own.
While using custodial solutions is convenient,
such solutions require Alice to entrust both her vault's key
and her assets to the same trusted party---%
introducing a single point of trust and
making users' assets vulnerable to breaches%
~\cite{certik2023poloniex,reuters2018coincheck,ussec2025cryptocustody}.
In contrast,
self-sovereign solutions do not depend on a single point of trust,
but typically require Alice to remember
(or securely store written copies of)
long alphanumeric strings,
such as passwords or mnemonic phrases%
~\cite{eleshin2025secrets}---%
imposing a memorability burden.

\emph{Social recovery}~\cite{
argent2021wallet,
loopring2023wallet,
indy,
derec2023recovery,
rathna2023social}
aims to address this trade-off
between custodial and self-sovereign solutions:
it enables
users to recover
a vault's key with
the help of trusted contacts, or \emph{trustees}.
Each trustee holds a \emph{share} of the vault's key,
and the user can recover the vault's key
by obtaining a threshold number of these shares.
Thus, social recovery eliminates
the need for memorizing long alphanumeric strings
without requiring a single point of trust.

Existing social recovery mechanisms, however,
impose an implicit memorability burden on users:
they expect users to remember
\emph{recovery metadata} (\cref{sec:problem}), such as
(i) the vault's existence;
(ii) how to invoke the recovery process;
(iii) configuration details such as the threshold;
and (iv) the identities of at least some trustees.
Since device loss is unpredictable,
and human memory naturally fades with time~\cite{murre2015replication},
we argue that it is undesirable and perhaps unreasonable
to expect users to memorize this data for
an indefinite period.
Some approaches
try to simplify memorability needs
by using a small, fixed number of trustees and threshold
(e.g., 3-of-5)~\cite{applerecovery,exodus}.
Imposing small, fixed parameters is inadequate, however:
(i) it fails to address the memorability issue,
as shown by Schechter et al.~\cite{schechter2009s},
where users struggled to recall \emph{3-of-4} trustees;
(ii) it increases the risk of collusion among
(or coercion of) the few trustees;
and
(iii) it increases user concerns about the suitability and
security of social recovery,
as users may hesitate to entrust recovery
to only a handful of contacts~\cite{mangipudi2025web3}.
We hold that
social key recovery should give users
control over their set of trustees and threshold,
while minimizing burdens of memorability.

We argue that social vault recovery should aspire to an ideal
we call \emph{self-recovery}:
a user should have some chance
of recovering a lost vault
even if she remembers \emph{nothing} about it,
not even its existence.
Suppose Alice loses her devices
and any memory of her vault---%
whether through accident-induced amnesia,
or simply because she used it rarely
for ``cold storage'' purposes and forgot it.
In time, Alice acquires new devices and,
driven by the natural human desire to stay
connected~\cite{baumeister2017need,holt2024social},
\emph{reconnects} with friends
who still hold shares of her old vault's key.
Through these reconnections alone,
her new device gathers enough shares
to \emph{rediscover} the vault
and recover it automatically.
While existing solutions burden Alice with
remembering metadata and managing recovery herself,
self-recovery integrates recovery into her natural
rebuilding process---allowing even her contacts to initiate it.

A second key issue compounding the memorability challenge
is that recovery metadata is itself privacy-sensitive.
An adversary who learns Alice's configuration and trustee set
gains a conveniently short list of targets
to manipulate, bribe, or threaten
in order to steal Alice's secrets.
Trustees who know each other pose a risk of their own:
after a falling-out with Alice,
a former friend could turn a threshold of trustees
against her---%
similar to documented patterns of
retaliatory behavior following interpersonal fallout%
~\cite{lalot2023unkindest,buckley2004reactions}.
The metadata is socially sensitive, too:
a close friend may be offended to learn
she was \emph{not} chosen as a trustee,
and revoking a no-longer-trusted friend's trusteeship
carries the awkwardness of ``unfriending''
on social media~\cite{gashi2016unfriending,lopez2013look}.
Even the mere fact that Alice has a vault
may attract adversaries' attention.
We therefore argue that social recovery solutions
should offer strong \emph{metadata privacy}:
no one but Alice should ever need to know
her recovery metadata---%
ideally not even that she has a vault.

Recovery metadata privacy
might, in principle, be achieved by asking Alice
to just remember the metadata and not entrust it to anyone.
However, this approach sacrifices the ideal of
self-recovery,
illustrating the tension between these two goals.
Other na\"ive solutions,
such as encrypting the recovery metadata
before asking Alice to store it in a safe place,
recursively create another key/vault recovery problem
with further memorability and metadata-privacy hazards.
We therefore pose this research question:
\begin{center}
    \emph{
	Can social vault recovery minimize
    memorability expectations by offering self-recovery,
	while protecting recovery metadata privacy?}
\end{center}

\begin{figure}[!t]
    \centering
        \includegraphics[width=0.45\textwidth]
        {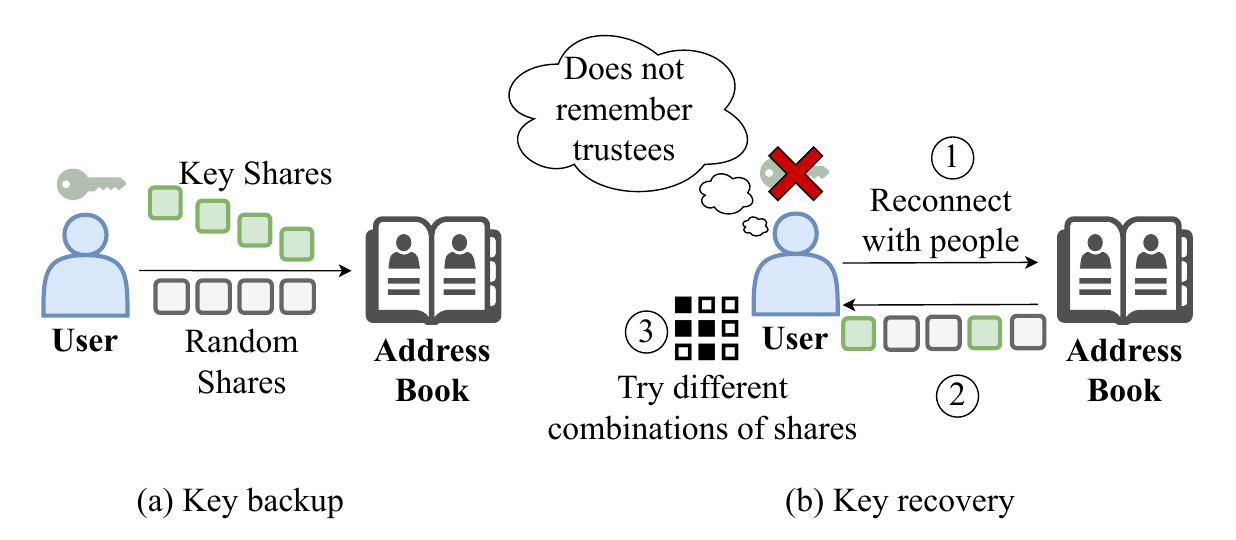}%
        \caption{Overview of \name.}
        \label{fig:overview}
\end{figure}

This paper presents \name,
a novel social key recovery framework
designed to address this challenging combination of requirements.
\cref{fig:overview} depicts \name's high-level design,
which is based on two key observations:
(i) after a loss, it is human nature to ``rebuild one's life''
including social circle~\cite{baumeister2017need,holt2024social};
and
(ii) if an adversary does not know the exact trustees set,
then she is forced to compromise more people
hoping to find enough shares for recovery eventually---%
which makes malicious recovery harder.
\name hides, or effectively anonymizes, Alice's chosen trustees
among Alice's broader circle of contacts
by distributing genuine recovery shares to trustees
while distributing cryptographically indistinguishable
random \emph{chaff} shares to all non-trustee contacts.
After device loss,
the natural process of socially re-connecting with prior contacts
gives the user's new Apollo-enabled device(s) access to a growing set of
the old (real or chaff) shares still held by her old contacts.
After each such reconnection,
Apollo automatically searches for combinations of shares
sufficient to recover any lost vault.
With a sufficient number of such reconnections, therefore---%
whether initiated by the user or by her contacts---%
Apollo eventually finds a threshold of shares
and uses them to recover vault access automatically,
without requiring any user-initiated recovery process.

There is always a risk that an adversary could find
and
compromise
enough trustees to steal Alice's vault.
However, an adversary in \name needs to obtain enough shares
without knowing who the trustees are and
without getting caught during the process.
In particular,
distributing indistinguishable real and chaff shares increases
the number of potential trustees,
forcing an adversary to compromise more people
to get enough relevant data.
This raises both the adversary's cumulative effort
and the chances of being caught,
and thus malicious recovery becomes less feasible.
To conceal even the vault's \emph{existence},
each contact's device must hold such shares regardless of
whether its owner uses a vault.
\name therefore targets an OS-integrated deployment,
though a standalone app is also possible
at some cost to privacy and self-recovery
(\cref{sec:disc:deployment}).

While indistinguishability provides self-recovery,
metadata privacy, and a defense against
malicious recovery attempts,
it introduces computation overhead during recovery.
Since key shares and chaff shares are indistinguishable,
the recovery algorithm must brute-force all possible
share combinations,
which grows exponentially
since $n \choose r$ is exponential in $r$. 
We address this efficiency bottleneck by
introducing \emph{multi-layered secret sharing} (MLSS)
(\cref{sec:apollo:mlss}),
a novel approach that shifts from the typical
social key recovery paradigm of using one layer
of Shamir's secret shares~\cite{shamir1979share}.
Existing secret sharing schemes
have employed multiple layers for assigning a hierarchy of
trust to trustees
~\cite{traverso2016dynamic,tassa2007hierarchical,
farras2012ideal}.
In contrast,
\multiss serves an orthogonal purpose:
it is designed solely
to reduce the computational overhead of anonymizing trustees
among a larger set of contacts.
By restructuring
the shares into two layers in MLSS,
we sidestep the exponential growth issue,
keeping the number of possible combinations
small and manageable
without compromising self-recovery or metadata privacy.

We implemented \name in Go~\cite{meyerson2014go}
and evaluated its performance.
After reconnecting with 30 people,
\name takes only 0.12 seconds to search over all combinations,
while the single-layered approach takes more than 25 hours.
In addition, we empirically analyze \name's security
against an adversary trying to recover by
wrongfully obtaining shares from contacts.
Even for a strong adversary
who can compromise every $1$-in-$2$ contacts,
Apollo reduces
the likelihood of malicious recovery to less than $0.1\%$.
While not cryptographically negligible,
this probability is low even under large-scale contact compromise
and can be further reduced by changing system parameters
(\cref{sec:eval:security}).

This paper makes the following novel contributions:
\begin{itemize}[leftmargin=*]
    \item A taxonomy of
        \emph{recovery metadata} in social recovery
        that elicits the need for \emph{self-recovery}
        and \emph{metadata privacy}
        (\cref{sec:problem}).
    \item The design of \emph{\name},
        the first social recovery framework
        to enable self-recovery with metadata privacy
        (\cref{sec:overview,sec:apollo}).
    \item The \emph{multi-layered secret sharing (MLSS)} scheme
        that makes recovery practical
        by lowering computation overhead
        (\cref{sec:apollo:mlss}).
    \item A prototype implementation of \name,
        along with its performance and empirical security analysis
        (\cref{sec:eval}).
\end{itemize}

\noindent{\bf Non-goals.} \name does not address the following:
\begin{enumerate}[leftmargin=*]
    \item
        Secure storage, availability, and vault management.
        Refer to~\cite{keybase,keeper,icloud} for details on these
        aspects of vaults.
    \item
        Adversarial ability to extort
        the key from the user itself.
        This issue is orthogonal to \name,
        and has been extensively studied in e-voting
        literature~\cite{juels2005coercion,
        lueks2020voteagain,
        estaji2020revisiting,
        merino2024vote,merino2025trip}.
    \item
        The integrity of backed-up shares:
        an adversary who compromises trustees can tamper with
        shares distributed.
        We discuss this attack in detail in \cref{sec:disc:integrity}.
\end{enumerate}

\section{Recovery Metadata Management Problem}
\label{sec:problem}
Consider Alice, a privacy-conscious smartphone user,
who wishes to set up an e2ee vault.
She decides against using custodial and self-sovereign solutions
since 
they rely on a single point of trust and 
impose a memorability burden,
respectively.
She considers social recovery approaches
that promise to minimize memorability burden
by enabling vault recovery with the help of
Alice's trusted contacts, or \emph{trustees}.
However, she realizes that
prior solutions impose a (hidden)
memorability burden by forcing Alice
to remember recovery-related information,
which we refer to as \emph{recovery metadata}.
For instance,
Alice needs to remember her trustees and the threshold
for recovering her vault.
Some prior solutions
potentially attempt to address this memorability issue by fixing
the system parameters (e.g., trustee count);
we review them next.

\begin{table}[!thbp]
    \centering
    \small
    \setlength{\tabcolsep}{2.5pt}
    \renewcommand{\arraystretch}{0.95}
    \begin{tabular}{lcccccc}
        \toprule
        & & \multicolumn{4}{c}{\textbf{Forgetfulness Tolerance}} & \\
        \cmidrule(lr){3-6}
        {\textbf{Protocol}} &
        {\textbf{Control}} &
        {\textbf{Exist.}} &
        {\textbf{Proc.}} &
        {\textbf{Trustees}} &
        {\textbf{Thresh.}} &
        {\textbf{Privacy}} \\
        \midrule
        Apple$^\text{\bf \dag}$~\cite{applerecovery} &
        \xmark &
        \xmark &
        \xmark &
        \xmark &
        \cmark &
        \xmark \\
        HTC$^\text{\bf \dag}$~\cite{exodus} &
        \xmark &
        \xmark &
        \xmark &
        \xmark &
        \cmark &
        \xmark \\
        Argent*~\cite{argent2021wallet} &
        \xmark &
        \xmark &
        \xmark &
        \xmark &
        \cmark &
        \xmark \\
        Loopring*~\cite{loopring2023wallet} &
        \xmark &
        \xmark &
        \xmark &
        \xmark &
        \cmark &
        \xmark \\
        DeRec~\cite{derec2023recovery} &
        \cmark &
        \xmark &
        \xmark &
        \xmark &
        \xmark &
        \xmark \\
        Indy~\cite{indy} &
        \cmark &
        \xmark &
        \xmark &
        \xmark &
        \xmark &
        \xmark \\
        Soc. Wallet~\cite{rathna2023social} &
        \cmark &
        \xmark &
        \xmark &
        \xmark &
        \xmark &
        \xmark \\
        \textbf{\name} &
        \cmark &
        \cmark &
        \cmark &
        \cmark &
        \cmark &
        \cmark \\
        \midrule[\heavyrulewidth]
        \multicolumn{7}{l}{\footnotesize
            * - fixed percentage threshold of $>50$;
            \dag - fixed configuration (e.g., 3-of-5).
        } \\
        \bottomrule
    \end{tabular}
    \caption{
        Table depicting
        if the memorability and privacy of
        recovery metadata is considered
        in existing solutions; users are expected
        to remember almost the entire metadata
        and none of them hide the
        trustee privacy.
    }
    \label{tab:comparison}
\end{table}

\subsection{The Memorability--User Control Trade-off}
\label{sec:problem:control}
The recovery configuration---%
\emph{who} and \emph{how many} must guard the key---%
is a central security decision of social recovery,
and it is inherently personal:
trust is distributed unevenly across every user's social circle.
A fixed configuration therefore does not fit all users:
a user with only three trustworthy contacts
must enlist weakly-trusted ones to fill a 3-of-5 quorum,
whereas a user with many trustworthy contacts
cannot opt for a larger, safer threshold.
Hard-coding this configuration thus does not remove a burden;
rather it makes a security decision on users behalf,
reinstating the system-dictated trust
that self-sovereign solutions set out to eliminate.
User control is thus a prerequisite for
self-sovereign recovery~\cite{allen2016path}.

Yet the first four solutions in \cref{tab:comparison}
hard-code this configuration.
Apple's recovery contacts~\cite{applerecovery}
and HTC Exodus~\cite{exodus}
require inputs from 1-of-5 and 3-of-5 trusted contacts for recovery,
respectively.
In contrast, Argent~\cite{argent2021wallet} and
Loopring~\cite{loopring2023wallet} do not fix
the trustees count but require shares from a majority for recovery.
This design, however, is problematic:
a small trustee set with a low threshold
raises the risk of collusion among
(or coercion of) trustees~\cite{buterin2021social},
and fixed parameters, being public knowledge,
tell an adversary exactly how many people to compromise.
Beyond security,
denying users this choice undermines
their autonomy and motivation~\cite{bennett2023does,peters2018designing},
and leads to more dissatisfaction
than an overload of choices~\cite{reutskaja2022choice}---%
a concern that only grows as
governments deploy digital-identity wallets
that place key management
directly in users' hands~\cite{euroeid,swisseid}.
Users must thus retain control over their trustee set
and threshold, though, as we show next,
this stands in tension with memorability.

\subsection{The Memorability Issue}
\label{sec:problem:memorability}
We next describe the four main recovery metadata
that prior solutions expect users to remember,
and analyze the drawbacks of this unrealistic expectation.

\noindent\emph{RM1: Vault Existence.}
All existing solutions expect Alice to remember
the fact that she had set up a vault.
However, this assumption is far from ideal
as an average smartphone user, like Alice,
installs numerous applications~\cite{snyderapps},
making it likely that she will eventually
forget about the vault application~\cite{googleapps}.
\smallskip

\noindent\emph{RM2: Recovery Procedure.}
Secondly, all existing solutions require
Alice to remember how she can recover her vault:
she needs to obtain shares
from the people she used to trust while backing up
her key.
Alice needs to remember this procedure
for an indefinite period
since she cannot foresee when she will
lose her device.
Since people's memory naturally fades with
time~\cite{murre2015replication},
it is unreasonable to expect users
to remember the recovery procedure,
as depicted
in the studies on fallback authentication%
~\cite{lassak2024comparative,micallef2021understanding}.
\smallskip

\noindent\emph{RM3: Trustee Details.}
Thirdly, Alice needs to remember the
trustee set.
Existing social key recovery mechanisms either:
(i) expect Alice to
``safely'' store the details of any number of trustees;
or (ii) fix a small number of trustees
(e.g., five recovery contacts in Apple).
The former design is inadequate as
Alice can
lose or forget trustees,
while the latter undermines user control
(\cref{sec:problem:control}).
\smallskip

\noindent\emph{RM4: Threshold.}
Finally, Alice needs to
remember additional configuration details,
such as the recovery threshold.
Since it would be difficult for a user to
remember the threshold
for an indefinite period~\cite{murre2015replication},
some solutions fix the threshold
(e.g., one in Apple's recovery contacts),
but they compromise on user control.
\smallskip

As \cref{tab:comparison} shows,
existing solutions burden users with remembering
almost the whole of the recovery metadata.
Some of them attempt to reduce
the burden either with
a fixed percentage threshold or
with a fixed configuration;
however, fixing parameters reduces user control,
which has major drawbacks (\cref{sec:problem:control}).
In summary, no existing solution
considers the consequences of users forgetting
the metadata while providing control to users.
With minor memorability issues,
recovery using existing solutions would
become an arduous task for Alice.

\subsection{Self-recovery}
\label{sec:problem:self}
We argue that social recovery solutions
must strive for the ideal of \emph{self-recovery},
such that Alice could regain access to the vault
despite not remembering any metadata.
To facilitate this ideal,
the recovery process must be embedded
within the broader context of
rebuilding one’s life after loss.
To this end, we leverage a key aspect of the rebuilding phase:
the intrinsic human need for social connection%
~\cite{baumeister2017need,holt2024social}.
This drive is evident in numerous real-world cases
where individuals reconnect with loved ones after
separation caused by wars or natural disasters%
~\cite{rescue2023how,people2025teen,people2025woman}.
Thus, by facilitating (self-)recovery via
Alice's and her contacts' natural motivation to reconnect,
Alice will eventually encounter enough people holding
her key shares,
without having to actively recall or seek them.
On the other hand, enabling self-recovery by trying to make
the metadata accessible, despite device loss, is ineffective:
storing a written copy of the
metadata in a secret location would only create a recursive problem,
forcing Alice to memorize how to access the secret location;
whereas, making the metadata public for accessibility
would present the exact route
for accessing Alice's assets to unwanted parties.
Moreover, if Charlie, one of Alice's contacts, learns that Alice did not
choose him as a trustee,
then it could strain the relationship
between Alice and Charlie.
Thus, there is a hidden privacy risk associated
with recovery metadata.

\subsection{Recovery Metadata Privacy}
\label{sec:problem:privacy}
Recovery metadata privacy implies that
no one, other than Alice, learns anything about
the recovery metadata---%
ideally, not even the vault's existence.
Prior solutions overlook the need to protect recovery metadata.
They fail to account for the risk
that an adversary could recover
the vault by compromising a subset of trustees,
leaving users responsible for selecting a sufficiently secure trustee set.
Hence, we next describe the reasons why
maintaining recovery metadata privacy is a must.
\smallskip

\noindent\emph{RM.PR1: Security.}
Leakage of any recovery metadata
could help an
adversary with malicious recovery:
(i)
the mere knowledge of
the vault's existence would attract adversaries
and the existence could
reveal additional information about the vault,
such as the vault's version,
which could be exploited
for finding vulnerabilities~\cite{nikitin2019reducing,scarlata2026zero}
(e.g., issues with
the encryption algorithm of a particular version).
(ii) the recovery procedure would tell an adversary how
to proceed for accessing Alice's vault;
(iii) trustee details would expose Alice’s exact social
closeness and inform an adversary which
individuals to target;
(iv) the knowledge of the threshold would give an estimate of the effort
needed for recovery.
\smallskip

\noindent\emph{RM.PR2: Preserving interpersonal relationships.}
Revealing whether someone is or isn't a trustee
may jeopardize Alice's relationship with her contacts.
This leakage is sensitive~\cite{yu2024dont}
because non-trustees might be disappointed to learn that Alice
did not consider them \emph{close enough}%
~\cite{jaffe2023secretive}.
Similarly, trustees might be disappointed
if Alice sets a high threshold while backing up
shares with them.
\smallskip

\noindent\emph{RM.PR3: User control.}
To avoid social awkwardness,
Alice may opt for
a socially acceptable set of trustees~\cite{cialdini2004social};
however, this choice undermines her control over her backup.
Alice might also fear being judged for
reclassifying trustees as non-trustees (or vice-versa),
limiting her ability to adjust her trustee set as she sees fit,
and hence suffering from the same issues highlighted earlier
(\cref{sec:problem:control}).

\paragraph{\bf Is recovery metadata privacy easily achievable?}
\label{sec:problem:privacy:soln}
Achieving only recovery metadata privacy seems simple:
have the user memorize the entire metadata,
and not store it elsewhere
(like prior solutions in \cref{tab:comparison}).
However, this naive design is inadequate for three main reasons:
(i) users need to memorize the metadata
for an unknown length of time;
(ii) the fear of forgetting trustees
may lead users to select fewer of them, akin to
users choose weak passwords to avoid forgetting strong ones%
~\cite{bonneau2012science};
(iii) if a user changes the trustees set,
the old trustees would be slighted to learn
that they are not trusted enough.
The third reason brings out a key property for
metadata privacy:
no one, other than the user, should know who the trustees are.
We refer to this property as \emph{trustee privacy},
and none of the existing solutions (\cref{tab:comparison})
achieves this property.

\begin{table}[h!]
\centering
\begin{tabular}{p{3.0cm} p{1.2cm} p{1.00cm} p{0.9cm} p{0.9cm}}
\toprule
\textbf{Solution} &
\textbf{\%trustees recalled} &
\textbf{Exp. recon.} &
\textbf{User Control} &
\textbf{Trust Status Leak} \\
\midrule
{\textbf{NS1}: User chooses} &
\multicolumn{1}{c}{
    \cellcolor{Green!20}100\%
} &
\multicolumn{1}{c}{
    \cellcolor{Green!20}10
} &
\multicolumn{1}{c}{
    \cellcolor{Green!20}Yes
} &
\multicolumn{1}{c}{
    \cellcolor{Red!20}20
} \\
trustees and asks&
\multicolumn{1}{c}{
    \cellcolor{Goldenrod!20}40\%
}&
\multicolumn{1}{c}{
    \cellcolor{Goldenrod!20}30
}&
\multicolumn{1}{c}{
    \cellcolor{Green!20}Yes
} &
\multicolumn{1}{c}{
    \cellcolor{Red!40}40
} \\ 
contacts she recalls&
\multicolumn{1}{c}{
    \cellcolor{YellowOrange!20}0\%
}&
\multicolumn{1}{c}{
    \cellcolor{YellowOrange!20}71
}&
\multicolumn{1}{c}{
    \cellcolor{Green!20}Yes
} &
\multicolumn{1}{c}{
    \cellcolor{Red!60}81
} \\ 
\midrule
\textbf{NS2}: Trustees chosen &
\multicolumn{1}{c}{
    \cellcolor{Green!20}100\%
}&
\multicolumn{1}{c}{
    \cellcolor{Green!20}10
}&
\multicolumn{1}{c}{
    \cellcolor{Red!40}No
} &
\multicolumn{1}{c}{
    \cellcolor{Green!20}0
} \\
randomly and user asks&
\multicolumn{1}{c}{
    \cellcolor{Goldenrod!20}40\%
}&
\multicolumn{1}{c}{
    \cellcolor{Goldenrod!20}30
}&
\multicolumn{1}{c}{
    \cellcolor{Red!40}No
} &
\multicolumn{1}{c}{
    \cellcolor{Green!20}0
} \\
contacts she recalls&
\multicolumn{1}{c}{
\cellcolor{YellowOrange!20}0\%
}&
\multicolumn{1}{c}{
    \cellcolor{YellowOrange!20}71
}&
\multicolumn{1}{c}{
    \cellcolor{Red!40}No
} &
\multicolumn{1}{c}{
    \cellcolor{Green!20}0
} \\
\midrule
\textbf{NS3}: Make all contacts&
\multicolumn{1}{c}{
    \cellcolor{Green!20}100\%
}&
\multicolumn{1}{c}{
    \cellcolor{RedOrange!20}75
}&
\multicolumn{1}{c}{
    \cellcolor{Red!40}No
} &
\multicolumn{1}{c}{
    \cellcolor{Green!20}0
} \\
trustees and user asks&
\multicolumn{1}{c}{
    \cellcolor{Goldenrod!20}40\%
}&
\multicolumn{1}{c}{
    \cellcolor{RedOrange!20}75
}&
\multicolumn{1}{c}{
    \cellcolor{Red!40}No
} &
\multicolumn{1}{c}{
    \cellcolor{Green!20}0
} \\
contacts she recalls&
\multicolumn{1}{c}{
    \cellcolor{YellowOrange!20}0\%
}&
\multicolumn{1}{c}{
\cellcolor{RedOrange!20}75
}&
\multicolumn{1}{c}{
    \cellcolor{Red!40}No
} &
\multicolumn{1}{c}{
    \cellcolor{Green!20}0
} \\
\midrule
{\textbf{Self-recovery +}} &
\multicolumn{1}{c}{
    \cellcolor{Green!20}100\%
}&
\multicolumn{1}{c}{
    \cellcolor{Green!20}10
}&
\multicolumn{1}{c}{
    \cellcolor{Green!20}Yes
} &
\multicolumn{1}{c}{
    \cellcolor{Green!20}0
} \\
\textbf{Metadata privacy}
&
\multicolumn{1}{c}{
    \cellcolor{Goldenrod!20}40\%
}&
\multicolumn{1}{c}{
    \cellcolor{Goldenrod!20}30
}&
\multicolumn{1}{c}{
    \cellcolor{Green!20}Yes
} &
\multicolumn{1}{c}{
    \cellcolor{Green!20}0
} \\
\textbf{(Apollo)}
&
\multicolumn{1}{c}{
\cellcolor{YellowOrange!20}0\%
}&
\multicolumn{1}{c}{
\cellcolor{YellowOrange!20}71
}&
\multicolumn{1}{c}{
    \cellcolor{Green!20}Yes
} &
\multicolumn{1}{c}{
    \cellcolor{Green!20}0
} \\
\bottomrule
\end{tabular}
\caption{
Comparison of na\"ive solutions
based on expected no. of reconnections,
user control,
and the no. of times trustee status is leaked for
$20$ trustees, $50\%$ threshold, $150$ contacts.
}
\label{tab:naive}
\end{table}

\subsection{Can na\"ive solutions work?}
\label{sec:problem:naive}
\cref{tab:naive} depicts why na\"ive solutions fail to achieve
user control, self-recovery, and metadata privacy together.

\noindent\emph{NS1: User approaches contacts hoping to find trustees.}
A simple solution is to use the existing protocols as is
and have Alice find trustees
by reaching out to her social circle.
This approach recovers the vault, but
does not protect trustee privacy, which could lead to
social awkwardness when non-trustees learn about their exclusion.

\noindent\emph{NS2: Choose trustees randomly.}
To avoid social awkwardness,
the recovery solution chooses the trustees randomly for Alice,
and she reconnects randomly in her social circle for recovery.
However,
having the application choose trustees on users' behalf
undermines user control,
which is a necessity
(\cref{sec:problem:control}).

\noindent\emph{NS3: Make everyone a trustee.}
An alternative approach is to designate all contacts as trustees,
allowing Alice to obtain a share from each known contact.
This design limits Alice's autonomy in trustee selection,
possibly imposing a higher threshold to reduce
the risk of collusion among less-trusted people.

In summary,
achieving self-recovery with metadata privacy
is a necessity and
na\"ively extending existing social key
recovery protocols does not suffice.
\name aims to address this issue.

\section{Overview of Apollo}\label{sec:overview}

This section covers
\name's system goals (\cref{sec:overview:goals}),
its high-level workflow
(\cref{sec:overview:workflow}),
its system model (\cref{sec:overview:system}),
and
its threat model (\cref{sec:overview:threat}).

\subsection{Goals}
\label{sec:overview:goals}

\noindent \name aims to achieve the following goals,
which are presented informally here
and made precise later
(\cref{sec:eval:security}).

\begin{itemize}[leftmargin=*]
    \item \emph{Low memorability expectations:}
    A user should be able to recover her key without
    remembering recovery metadata.
    \item \emph{Confidentiality:}
    The backup process should minimize leakage of
    recovery metadata to others, even to trustees.
    \item \emph{Unauthorized access mitigation:}
    An adversary,
    trying to recover a user's key by
    compromising contacts,
    should fail with high probability.
    \item \emph{Efficiency:}
    The recovery algorithm 
    should execute within a few minutes,
    even for large address books.
\end{itemize}

\subsection{Apollo's workflow}
\label{sec:overview:workflow}
\begin{figure}[htbp]
    \centering
        \includegraphics[width=0.495\textwidth]
        {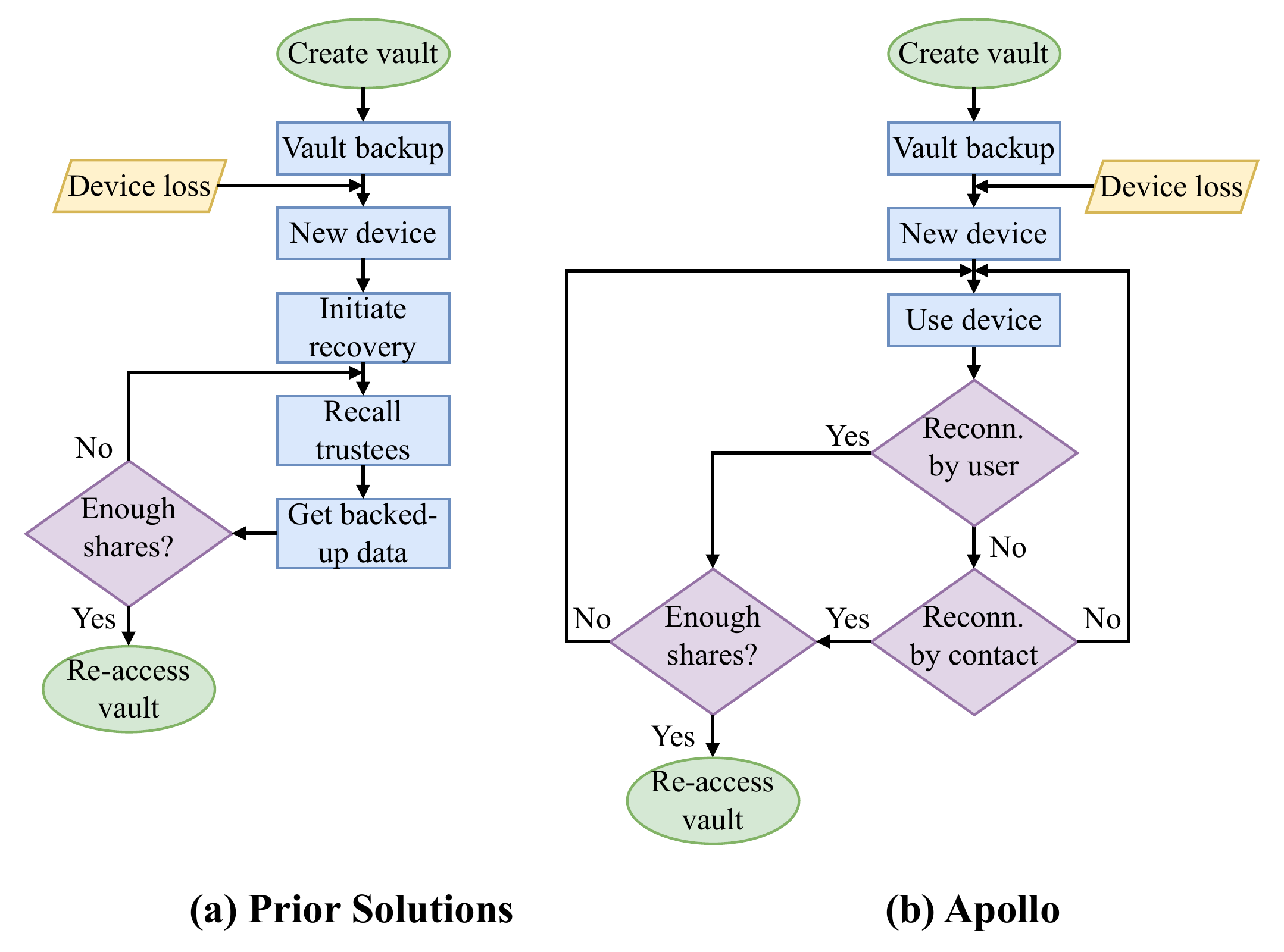}
        \caption{
            Workflow comparison:
            while existing solutions need a user to recall and
            actively seek her trustees,
            Apollo's recovery progresses with natural reconnections
            with lost contacts,
            which might be even initiated by user's contacts.
        }
        \label{fig:overview:workflow}
\end{figure}
\cref{fig:overview:workflow} contrasts \name's workflow
with that of prior solutions.
Prior solutions back up shares with the user's trustees
and leave recovery entirely to the user:
after device loss,
she must recall the vault and her trustees,
and actively seek them until she gathers enough shares.
Furthermore, she needs to remember all recovery metadata
indefinitely,
and this design risks social discomfort,
as mistakenly approaching a non-trustee
reveals their exclusion.

\name instead lets recovery proceed via \emph{reconnections},
which either the user or her contact(s) may initiate.
To this end,
\name distributes \emph{indistinguishable} real and chaff shares
to trustees and non-trustees, respectively,
making every contact a potential trustee.
Each reconnection thus yields a share (real or chaff)
while leaking neither trustee status nor any other metadata,
and as the user rebuilds her social life,
she eventually gathers enough real shares
to recover her vault---%
without ever recalling or seeking trustees.

\subsection{System and Communication Model}
\label{sec:overview:system}
\paragraph{\bf System Model}
There are four main components of \name:
(i) \emph{user};
(ii) \emph{vault};
(iii) \emph{vault's key};
and
(iv) \emph{address book} of the user (contacts).
Users and their contacts own a personal device
(e.g., smartphone) with the \name application installed.
The key is a bit string of variable length,
and we focus on the backup and recovery of the key;
we refer to~\cite{keybase,keeper,icloud} for other
aspects of vaults.
\name can be deployed as part of the OS
(e.g., App Store in iOS or Play Store in Android)
or as a separate application.
We consider the former ideal in \name's design
since it:
(a) eases onboarding
by eliminating the manual installation step; and
(b) enables stronger privacy properties,
which we detail
in \cref{sec:disc:deployment}.
However, \name is not a part of address book software
since users typically back up their address books with cloud services,
and we avoid storing key-related information in the cloud.
\name divides the address book into:
\emph{\trustees}, the people who hold relevant
recovery data, and
\emph{\nontrustees}.
\name uses Shamir's secret sharing~\cite{shamir1979share}
to generate key shares for trustees.
The user can manually
update her trustees, \nontrustees,
and the threshold for secret shares.

When a user loses her device and
thus, her vault's key,
she reconnects with her
contacts using a new device:
the user and her contacts exchange
and store each other's details
(e.g., phone number) on their devices.
The reconnection could be either initiated by the user,
who tries to maintain contact with the people she knows,
or by the contacts, who wish to stay in touch with her.
We assume that contacts' devices:
(i) store the backed up information,
(ii) are responsive;
and the user's new device
reliably obtains blobs
from these contacts after reconnecting.

\paragraph{\bf Communication Model}
We assume that the user's device
and the contacts' devices are connected over
an asynchronous network, allowing them
to communicate over an
end-to-end encrypted channel,
used for exchanging packets during key backup
and key recovery.
A user reconnects with her contacts
using an authenticated out-of-band channel
and proves her identity to them during recovery.
For example, a user can meet someone in-person
or connect
via a phone call with the help of
another friend.

\subsection{Threat model}\label{sec:overview:threat}
The adversary's goal in \name
is to reconstruct a user's key by obtaining
backed-up information from the user's contacts.
The adversary can
phish, impersonate, bribe, coerce,
or use any other means to compromise the user's contacts.
However, defining the precise number of contacts
that can be compromised in a social recovery setting,
capturing both diverse real-world corruption vectors
(e.g., phishing or coercion)
and user-specific trust dynamics,
is complex.
Thus, for clarity,
we initially consider a \emph{baseline adversary}
who does not know the trustees' identities and
compromises an arbitrary set of contacts.

The baseline adversary can compromise
any $f$-of-$\numaddrbook$ people holding shares
(user's contacts)---%
consistent with the adversarial models
in secret-sharing literature%
~\cite{shamir1979share,beimel2011secret}.
This model isolates the core mechanics of
\name and simplifies the presentation.
In \cref{sec:discussion},
we introduce a more realistic adversarial model,
and in \cref{sec:eval:security},
we demonstrate that \name's defenses
remain effective under both models.
We leave side channel attacks
out of the scope of this work.

\section{Apollo Protocol Design}
\label{sec:apollo}
\begin{figure*}[thpb]
\centering
    {\includegraphics[width=0.80\textwidth]
    {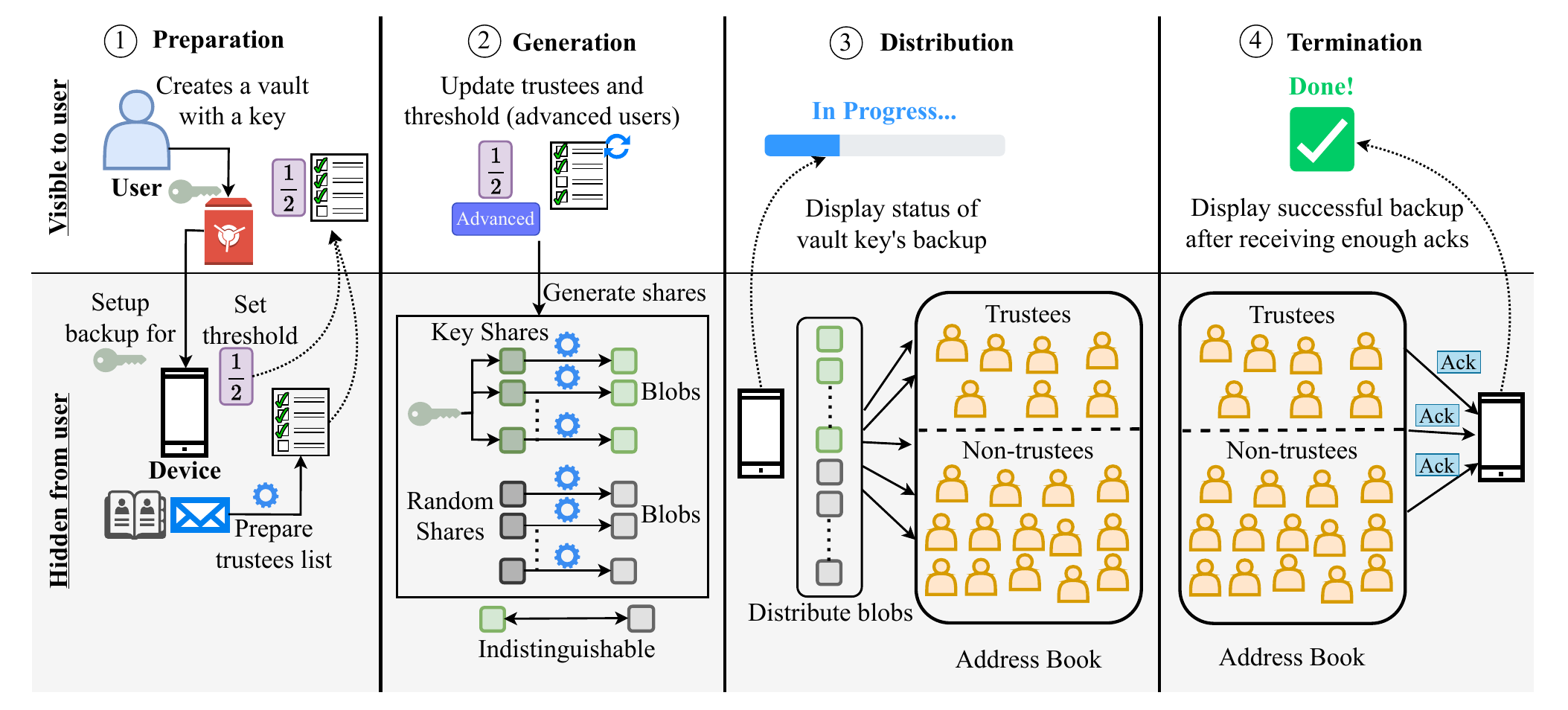}%
    \caption{Key backup: figure does not include
    devices of contacts for clarity.}
    \label{fig:anon-workflow:backup}}
\end{figure*}

This section details \name's design
for setting up a recovery mechanism
for Alice's e2ee vault,
ensuring both self-recovery and
recovery metadata privacy.
First, we explain the key backup (\cref{sec:apollo:backup})
and the key recovery process (\cref{sec:apollo:recovery}).
Then, we outline the multi-layered secret sharing
scheme (\cref{sec:apollo:mlss}).

\subsection{Backup}
\label{sec:apollo:backup}
\cref{fig:anon-workflow:backup}
illustrates the backup procedure.
The numbers in the figure show the sequential steps,
which we detail below.

\paragraph{\bf 1) Preparation}
The backup process starts with Alice
setting up a vault on her device.
After the vault is ready,
\name uses information
associated with the address book,
such as messages and call logs,
to recommend a tentative set of trustees.
The primary motivation behind recommending \trustees
is to mitigate the initial friction in \name's adoption
by alleviating the ``blank slate'' problem%
\cite{davidson18automatically}:
giving Alice a starting point to
choose her \trustees,
although she can edit this list based on her needs.
The remaining contacts form her \nontrustees.
To finish the preparation phase,
\name needs a threshold---the fraction of
\trustees needed for recovery.
To reduce the burden,
\name sets a default threshold,
although advanced users can
set this value themselves.

\paragraph{\bf 2) Generation}
After the preparation phase,
\name proceeds to generate shares.
\name uses Shamir's secret sharing~\cite{shamir1979share}
to generate shares of the key for \trustees
(\emph{key shares}), and 
it generates shares that are irrelevant to recovery
(\emph{chaff shares}) for \nontrustees.
After generating shares,
\name constructs an \emph{indistinguishable} blob,
containing a share and a \checksum.
The \checksum is essential for verifying
successful recovery ,and like random shares,
tags held by non-trustees
do not help with recovery.
A \nontrustee's blob comprises a \checksum for
maintaining indistinguishability solely.

\paragraph{\bf 3) Distribution}
Alice's device sends one or more blobs
to each contact's device.
On receiving blobs,
a device sends an acknowledgment back to Alice's device.
During this phase,
Alice's device could provide the backup's progress
based on the number of acknowledgments received.
Since \name is part of the OS,
the distribution phase runs in the background,
without needing Alice's or her contacts' active involvement.
In turn,
Alice does not need to bother or convince her contacts
to store random blob(s),
making the backup more convenient,
unlike existing approaches~\cite{yu2024dont}.

\paragraph{\bf 4) Termination}
After Alice's device receives acknowledgments
from enough contacts,
the backup ends,
and the device notifies Alice about the successful backup.
If Alice's device does not receive an acknowledgment
from a contact after a certain amount of time
(which could be a tunable parameter),
then it re-sends the corresponding blob(s).
For the core design of \name,
we assume that every contact will be online
at some point in the near future,
and thus,
Alice's device eventually receives all the acknowledgments.
We discuss the case when this assumption does not hold
in~\cref{sec:discussion}.

\subsection{(Self-)Recovery}
\label{sec:apollo:recovery}
\begin{figure}[htb]
    \centering
        \includegraphics[width=0.48\textwidth]
        {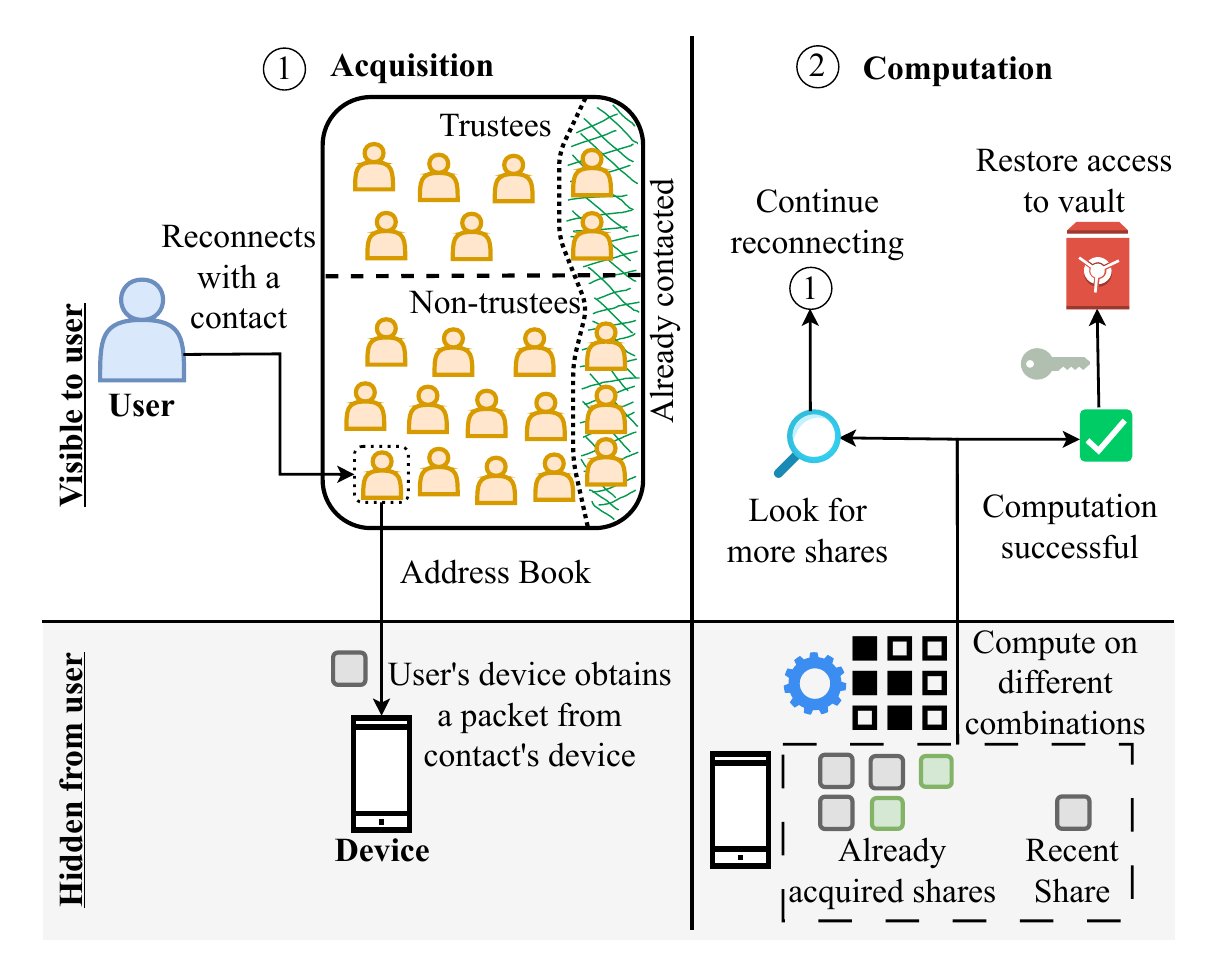}%
        \caption{Key recovery.}
        \label{fig:anon-workflow:recovery}
\end{figure}

We assume that Alice loses her device(s) and does not remember any of
the recovery metadata (\cref{sec:problem}).
She buys a new device to rebuild her life.
Over time, she \emph{reconnects} with her lost contacts,
either by reaching out herself or by responding when they contact her.
\name aims to support recovery through these
natural reconnection events,
even without Alice remembering any metadata or
even realizing that she is engaging in a recovery process.
\cref{fig:anon-workflow:recovery} depicts the key recovery process.
The numbers in the figure show the sequential recovery steps,
which we detail below.

\paragraph{\bf 1) Acquisition}
While reconnecting with each other,
Alice and her contact exchange contact information with each other.
Then, Alice's device requests the blob(s) from the
reconnected contact's device.
As Alice continues to reconnect with people,
her device acquires blob(s) and
runs computation on them.

\paragraph{\bf 2) Computation}
Alice's device tries different combinations of obtained shares
to reconstruct the key.
If no combination yields the key,
\name waits for Alice
to reconnect with the next person.
After another reconnection,
Alice's device tries combinations containing
shares obtained from the most recent blob(s).
After Alice reconnects with \emph{enough} trustees,
\name reconstructs the key and restores access to
the vault.
Therefore, the \name application,
which is part of the device's OS,
handles all the computation for vault recovery
in the background
while Alice solely focuses on rebuilding her life
and reconnecting with people.

\paragraph{\bf Self-recovery + metadata privacy}
Since \name is a part of the device OS,
Alice does not need to worry or even need to know
about the exchange of blobs and the computation.
She simply needs to reconnect with her
lost contacts, and \name handles the rest,
thus ensuring self-recovery (\cref{sec:problem:self}).
Additionally, trusteeship remains hidden
as each contact holds a random blob.
A blob does not contain any information
about a vault's existence, the recovery procedure,
or the threshold.
Therefore, recovery metadata privacy (\cref{sec:problem:privacy})
is preserved.

We next provide details on the
multi-layered secret sharing. 
As a warmup for \name's multi-layered approach,
we briefly explain the recovery using
a single layer of Shamir's secret sharing.

\subsection{Single-layered Approach}
\label{sec:apollo:single}

\begin{figure}[htb]
    \centering
    \begin{tikzpicture}
        \node[blobcell] (s) at (0,0) {share};
        \node[blobcell] (h) at (s.east) {$\hash{\key}$};
    \end{tikzpicture}
    \caption{Structure of a blob in the single-layered approach;
        $\mathsf{H}(\key)$ is the hash of the key.
    }\label{fig:blob-single}
\end{figure}

\paragraph{\bf Packet structure}
The single-layered approach
backs up one Shamir's secret share of the key, $\key$, per \trustee,
and stores an indistinguishable chaff share
with a \nontrustee.
The hash of the key, $\key$,
is concatenated with a share to create a blob,
and the hash acts as an indicator of successful reconstruction.
\cref{fig:blob-single} depicts the structure of this blob
received by each contact.
Assuming the random oracle model,
the blobs are indistinguishable, and
they do not leak any extraneous information,
making it impossible
to determine if
the contact is a trustee
or a non-trustee.

\paragraph{\bf Key Recovery}
After reconnecting with ${\n}_{\obt}$ people,
Alice's new device has a set of blobs
$B_{\obt} = \{\blob{1}, \dots, \blob{{\n}_{\obt}}\}$.
Let ${\addrbook}_{\obt} = \{\addrbookpart{1}, \dots,
\addrbookpart{{\n}_{\obt}}\}$ be
the set of people in the address book that
Alice has reconnected with,
$\threshold$ be the percentage threshold,
and $\trusteesset$ be the set of trustees.
Then, $B_{\obt}$ is fed to the reconstruction algorithm:

\begin{align*}
    \mathsf{Reconstruct}_{\single}({B_{\obt}}) =
    \begin{cases}
        \key & \text{if }
        | ({\addrbook}_{\obt} \cap \trusteesset) |
        \geq \lceil {\tau |\trusteesset|} \rceil \\
        \nullval & \text{otherwise}
    \end{cases}
\end{align*}

$\mathsf{{Reconstruct}}_{\single}$ tries all possible
combinations of shares in the blobs with
\emph{all possible thresholds}.
If the output is $\nullval$, the device
waits for the next reconnection.
Thus, $\mathsf{{Reconstruct}}_{\single}$ reconstructs the key
after Alice reconnects with a threshold number of trustees.
However,
this design is not scalable due to
the exponential and polynomial growth of the number of combinations with
the threshold
and the address book size, respectively
(we provide the bounds on this
growth in~\cref{app:computation-complexity}).
To make \name scalable,
we next provide the novel
\emph{Multi-layered Secret Sharing} scheme.

\subsection{Multi-layered Secret Sharing}
\label{sec:apollo:mlss}
\begin{figure}[htb]
  \centering
    \includegraphics[width=0.45\textwidth]
    {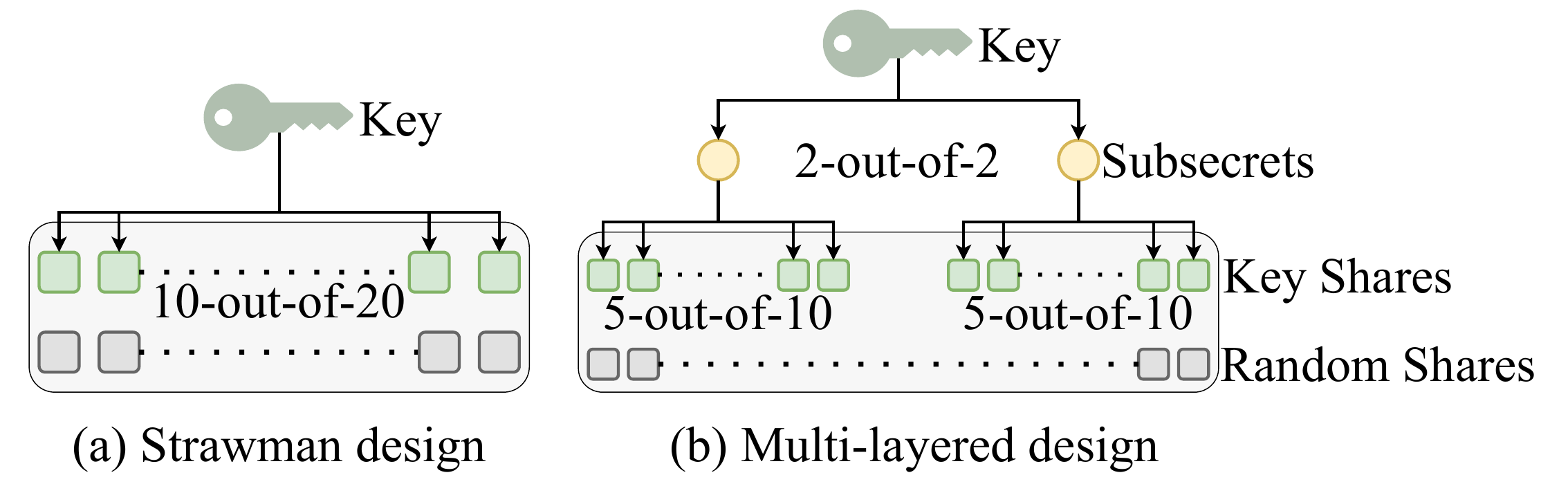}%
    \caption{Secret distribution structure.}\label{fig:multi-layered}
\end{figure}

Multi-layered secret sharing (MLSS)
addresses the scalability issue
of the single-layered approach
by structuring the shares of the key ($\key$) into two layers:
it first generates \emph{subsecrets},
the additive shares of $\key$, and
then generates Shamir's secret shares of
subsecrets (key shares).
The key shares are generated with a fixed threshold,
referred to as the \emph{absolute threshold} (${\absolute}$).
Each subsecret has
$\lfloor \frac{{\absolute}}{{\threshold}} \rfloor$ shares,
ensuring a threshold of at least $\tau$ for each subsecret.
MLSS distributes key shares among \trustees,
while it backs up chaff shares
among \nontrustees.
Chaff shares are indistinguishable from key shares
and are irrelevant to key recovery.
\cref{fig:multi-layered}
illustrates the share generation
in \multiss
compared to the single-layered approach
with $20$ trustees, ${\threshold} = 0.5$, and ${\absolute} = 5$.
Using additive shares for subsecrets
ensures that Alice must obtain \emph{at least}
the same fraction of shares as
the percentage threshold, ${\threshold}$.
Finally, since one needs to recover all the subsecrets,
\multiss can use a low threshold for shares of subsecrets ($\absolute$),
while having a high overall threshold for the key.

\paragraph{\bf Share distribution}
For each subsecret, $\share{l}$,
\multiss generates
$\sharesnum_{\sub} =
\lfloor {\absolute}/{\threshold} \rfloor$
shares with a threshold of $\absolute$.
In total, \name generates
$\sharesnum_{\totalnum} = \sharesnum_{\sub} \times {\numss}$ shares,
where $\numss$ is the number of subsecrets.
\name randomly samples without replacement
$\sharesnum_{\min} =
\lfloor {\sharesnum_{\totalnum}}/{|\trusteesset|} \rfloor$
shares for each trustee.
If
$\sharesnum_{\totalnum} \ (\mathsf{mod} \ |\trusteesset|) \neq 0$,
\name assigns the remaining shares
uniformly without replacement among some trustees,
distributing shares among trustees
(almost) uniformly.
A uniform distribution
is more secure as
no trustee can be more vulnerable than another.
We analyze this claim in \cref{app:security:uniform}.
For indistinguishability,
the number of shares per trustee must be the same, and hence,
\multiss assigns a random share to all trustees
who are assigned $\sharesnum_{\min}$ shares, while
each \nontrustee is assigned
$(\sharesnum_{\min}+1)$ random shares.
On the other hand,
when $\sharesnum_{\totalnum} \ (\mathsf{mod} \ |\trusteesset|) = 0$,
each trustee and \nontrustee stores $\sharesnum_{\min}$
key shares and chaff shares, respectively.
This share assignment mechanism ensures every contact holds
the same number of distinct shares,
ensuring indistinguishability.

\paragraph{\bf Blob}
\begin{figure}[htb]
  \centering
    \begin{tikzpicture}
        \node[blobcell] (s) at (0,0) {share};
        \node[blobcell] (t) at (s.east) {salt};
        \node[blobcell] (h1) at (t.east) {$\Hofstuff{\salt}{\mathrm{ss}}$};
        \node[blobcell] (h2) at (h1.east) {$\Hofstuff{\salt}{\key}$};
    \end{tikzpicture}
    \caption{Structure of a blob in MLSS; ss depicts the
    subsecret to which the share corresponds to;
    $\mathsf{H(.)}$ is the hash function.
    }
    \label{fig:blob-mlss}
\end{figure}

\multiss
assigns one or more shares to each contact, and
creates a blob for each share.
For successful recovery indication,
\multiss stores a \checksum in a blob.
The structure of the blob is depicted in~\cref{fig:blob-mlss}.
If we had used the blob structure of
the single-layered approach (\cref{sec:apollo:single}),
we would need to store all the subsecrets' hash and
the key hash with every contact for indistinguishability.
However, this design not only leaks the number of subsecrets
but also makes the recovery algorithm slower by making it check
with all the subsecrets.
To deal with this,
\multiss uses cryptographic salts for randomizing the \checksum:
a blob stores the salted hash of the relevant subsecret.
This prevents any leakage about the number of subsecrets and enables
the recovery algorithm to check with the relevant hash only.
For blobs containing random shares,
two (pseudo)random strings are added to ensure indistinguishability.

We formally analyze this blob structure in
\cref{app:security:formal}.
We define metadata privacy as a game in which
an adversary chooses two recovery configurations---%
the address book size, the number of trustees,
and the threshold, or ``no vault at all''---%
and must tell, from the blobs of a contact,
which configuration was used.
We prove in the random oracle model that
the adversary's advantage is negligible:
the blobs of any coalition of contacts
below the recovery threshold are indistinguishable
from pure chaff, a distribution that carries no metadata.
The analysis covers trustee, threshold, and
vault-existence privacy, and additionally shows that
the shares reveal nothing about $\key$
(information-theoretically) as long as
the coalition is below the threshold
of at least one subsecret.

\paragraph{\bf Number of subsecrets}
\label{sec:apollo:mlss:num-ss}
Fixing the number of subsecrets
would leak information about the threshold,
as the number of blobs would vary with threshold.
Hence,
\name fixes the number of shares held
by a contact ($\spp$).
Consequently,
we use a low $\spp$ which leads to fewer share combinations,
improving \name's performance.
Even after fixing $\spp$,
there could be
multiple possible values of
the number of subsecrets ($\numss$).
For example, with
${\absolute} = 3$, ${\threshold} = 0.5$,
$ |\trusteesset| = 20$,
consider two possible numbers of subsecrets:
$\numss = 5$ and $\numss = 6$.
For $\numss = 5$,
only $10$ \trustees hold $2$ key shares and the rest hold a key share
and a random share;
for $\numss = 6$,
$16$ trustees hold $2$ key shares and $4$ of them hold $1$.
$\numss = 6$ is better since:
(i) more subsecrets need to be recovered,
making malicious recovery harder;
(ii) distributing fewer or no random shares among \trustees
aligns \multiss's recovery profile
with that of a single-layered setting
(see \cref{app:security:multi:perfect}).

\paragraph{\bf Key Reconstruction}\label{sec:mlss:reconstruction}
\begin{algorithm}[htb]
    \caption{$\mathsf{Reconstruct}_{\multiss}$}\label{alg:reconstruction-mlss}
    \begin{algorithmic}[1]
    \State \textbf{Input}: $B_{\obt} = \{\blob{i}\}_{i=1}^{\spp\times \n_{\obt}}, \alpha$
    \State \textbf{Output}: $\key_{\rec} \in \{\key, \nullval\}$
    \State $K = \{\share{i}\} \gets \mathsf{ExtractShares}(B_{\obt})$
    \State $H \gets \mathsf{GetHashDict}(B_{\obt})$,
    $R \gets \mathsf{GetSaltDict}(B_{\obt})$
    \State $\key_{\rec} \gets \nullval, \recovered \gets \mathbf{false},
    S \gets [ \ ]$
    \State $C = \{\{\share{i}\}_{i=1}^{\alpha}\} \gets
    \mathsf{GetCombinations}(K, \alpha)$
    \For{$c \in C$}
        \State $\ssmath{\interp} \gets \mathsf{Interpolate}(c)$
        \State $H_{\relev} \gets H[c[0]], \salt \gets R[c[0]],
        \key_{\obt} \gets \nullval$
        \For{$h \in H_{\relev}$}
            \If{$\Hofstuff{\salt}{\ssmath{\interp}} = h
            \text{ and } \ssmath{\interp} \notin S$}
                \State $\mathsf{append}(S, \ssmath{\interp})$
                \State $\key_{\obt} =  \mathsf{Add}(S)$ \label{line:add}
            \EndIf
        \EndFor
        \If{$\key_{\obt} \neq \nullval \text{ and }
        \mathsf{len}(S) \neq 1$}
        \For{$h \in H_{\relev}$}
            \If{$\Hofstuff{\salt}{\key_{\obt}} = h$}
                \State $\key_{\rec} \gets \key_{\obt},
                \recovered \gets \mathbf{true}$
                \State \textbf{break}
            \EndIf
        \EndFor
        \EndIf
        \If{$\recovered = \mathbf{true}$}
            \State \textbf{break}
        \EndIf
    \EndFor
    \State \textbf{return} $\key_{\rec}$
    \end{algorithmic}
\end{algorithm}

After contacting $n_{\obt}$ contacts,
the obtained blobs,
$B_{obt} = \{\blob{1}, ...,\blob{\spp\times n_{obt}}\}$,
are fed to the reconstruction
algorithm of \name,
$\mathsf{Reconstruct}_{\multiss}$, which is depicted
in~\cref{alg:reconstruction-mlss}.
$\mathsf{Reconstruct}_{\multiss}$
generates combinations of ${\absolute}$ shares
and tries reconstructing subsecrets from them.
A hash match indicates the recovery of a subsecret, and
$\mathsf{Reconstruct}_{\multiss}$ tries to
reconstruct $\key$ by adding the recovered subsecrets.

\begin{figure}[htb]
    \centering
        \includegraphics[width=0.34\textwidth]
        {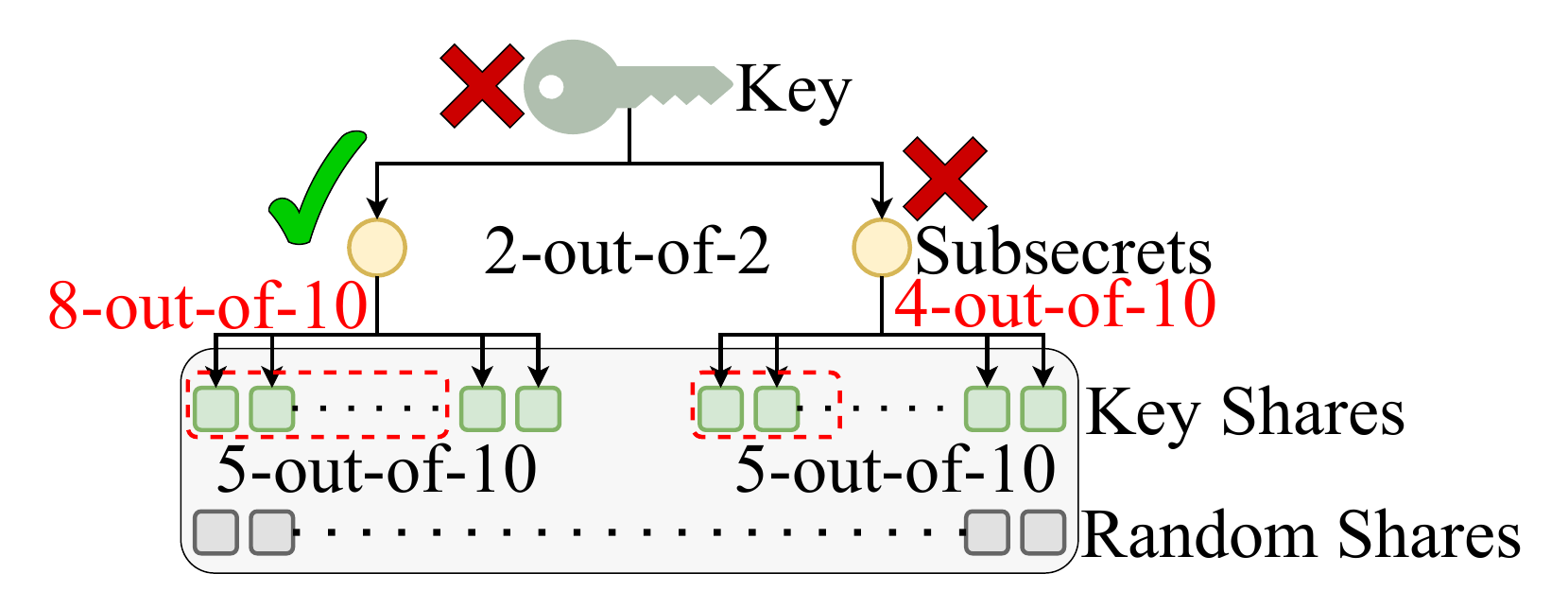}
        \caption{Non-recovery of key despite exceeding threshold
        as enough shares have not been obtained for a subsecret.}
        \label{fig:multi-layered-fail}
\end{figure}

\paragraph{\bf Probabilistic Recovery}
\label{sec:apollo:mlss:prob}
Although \multiss ensures efficient recovery,
it might require a user to connect with
more than threshold trustees.
Suppose a user reconnects with $12$ \trustees
and obtains shares as shown in~\cref{fig:multi-layered-fail}.
Despite more than threshold trustee reconnections,
\multiss fails to reconstruct $\key$
as one of the subsecrets is not recovered.
While this property of \multiss may require more than
threshold reconnections,
it also increases the effort required from an adversary.
\cref{app:practical:variation} provides an
analysis of the deviation of \multiss from
the single-layered approach,
and \cref{app:practical:variants} illustrates
three variants that increase
recovery probability at the user-defined threshold.
There is also a small probability of recovery with less
than threshold reconnections,
but this happens with very low probability,
as we show in \cref{sec:eval:security};
in \cref{app:security:multi:perfect},
we show how we can make this probability zero.

\subsection{Security against malicious recovery}
Since \name is a part of the OS,
every contact uses \name by default.
Thus, each of these contact stores
an indistinguishable blob,
which makes a user's entire address book
an anonymity set for the trustees.
Therefore, the baseline adversary (\cref{sec:overview:threat}),
who does not know trustees,
has no choice but to compromise an arbitrary set of $f$ contacts.
Since this adversary compromises contacts
``in the hopes of getting key shares'',
the adversary does not have a high probability of recovering
the user's key unless she manages to compromise
a large fraction of contacts, which makes
malicious recovery less feasible.
In \cref{sec:eval:security},
we back up this claim with empirical analysis,
and also show why malicious recovery gets harder in practice
when we evaluate a realistic adversary.

\subsection{Achieving the System Goals}
We summarize
how \name achieves the
system goals (\cref{sec:overview:goals}).
\begin{itemize}[leftmargin=*]
    \item \emph{Low memorability.}
    With \name,
    even if Alice forgets the recovery metadata,
    it does not impede recovery;
    she can re-access her vault
    as long as she continues to reconnect,
    which will be a natural part of Alice rebuilding her life.
    \item \emph{Confidentiality.}
    Since the blobs in \name are indistinguishable
    (see \cref{app:security:formal} for game-based proofs),
    no recovery metadata is leaked,
    and no one can detect if the blob holds relevant data,
    which in turn protects trustee
    privacy.
    \item \emph{Unauthorized access mitigation.}
    Due to the inclusion of \nontrustees,
    the \trustees are hidden within a larger population and hence,
    the adversary has to exert more effort
    as she would obtain chaff blobs from \nontrustees.
    This makes malicious recovery a daunting task, and
    we evaluate this gain in security
    in~\cref{sec:eval:security}.
    \item \emph{Efficiency.}
    The multi-layered approach generates
    combinations with a small fixed threshold,
    which bypasses the exponential growth.
    We depict the performance gain
    in~\cref{sec:eval:practical,sec:eval:overhead}.
\end{itemize}

\section{Discussion}
\label{sec:discussion}

\paragraph{\bf Realistic adversary}
\label{sec:disc:realistic}
Using the standard $f$-of-$n$ adversary model does not cover all
the aspects of a realistic adversary.
Firstly, since key shares are digital secrets,
and secrets are closely tied to
interpersonal trust and relationship maintenance%
~\cite{jaffe2023secretive,bedrov2024just,
fiske2018social,canary2015relationship},
contacts will not trivially disclose
their shares to an adversary.
Moreover,
technical and physical constraints may limit
the adversary---for instance,
phishing a tech-savvy contact or coercing one abroad.
An adversary might even
get reported while trying to compromise contacts---%
consistent with evidence that
users report phishing attacks~\cite{kersten2022investigating,lain2022phishing}
and that bystanders intervene when witnessing harm to
known contacts~\cite{davidovic2023intervene,levine2008responsive}.
Our realistic model also accounts for users forgetting their trustees,
and we base our analysis on Ebbinghaus' curve~\cite{murre2015replication}.
While Ebbinghaus' curve does
capture memory decay,
it concerns syllable recall---%
not recalling trusted contacts.
To better understant the ability to recall trustees,
we need a usability study for \name,
which we leave for future work.
Moreover, since others typically misjudge
a person's knowledge of themselves~\cite{vazire2010knows,pronin2001you}, 
we assume an adversary knows
fewer trustees than the user recalls.
We analyze these uncertainties in detail
and depict \name's defense against
realistic adversaries in \cref{sec:eval:security}.

\paragraph{\bf Deployment scenarios}
\label{sec:disc:deployment}
Although we envision \name to be a part of the OS,
\name can be a separate application integrated with the address book.
However, when used as a separate application,
only interested users would install \name
on their devices,
leaking the vault’s existence to an adversary.
In contrast,
when part of the OS,
\name can store a random blob for each
contact, irrespective of the contact owning a vault,
thereby keeping vault's existence private.
A standalone deployment also weakens self-recovery:
the user must remember her vault's existence (\emph{RM1})
to reinstall \name after device loss.
She still need not recall the procedure, trustees,
or threshold (\emph{RM2--RM4}),
keeping \name less demanding than
prior solutions (\cref{tab:comparison}).
Furthermore, since users often communicate
across multiple applications~\cite{datareportal},
default call and SMS records may not reflect true interaction patterns;
ideally, trustee recommendation should use
statistics from multiple applications.

\paragraph{\bf Non-responsive contacts}
The backup in \name's main design would stall
if some contacts are offline
since the backup waits for
acknowledgments from \emph{all} contacts.
To address this issue,
the backup could finish after receiving
acknowledgments from a majority of \trustees and contacts.
Thresholds must be defined for both \trustees and contacts
since acknowledgments from sufficient contacts
do not imply sufficient trustee acknowledgments,
and vice versa.
Finally, some \trustees may lose
touch over time
and hence, might not during recovery.
To counter this,
the user's device can periodically send heartbeat signals
and prompt her to update the \trustees
if a trustee is unreachable
for an extended period.

\paragraph{\bf Transient relationships}
Since trust varies with time,
\name must allow the update of configuration:
adding new contacts, changing contact status
(e.g., a trustee to non-trustee),
or changing the threshold.
\name would need to redistribute all shares
during such configuration updates.
In addition, transient trust can pose
a security issue: if a user falls out with a threshold of trustees,
those trustees could collude to reconstruct the key.
This risk persists unless old shares are invalidated.
This can be done by
changing the vault's key and distributing fresh shares,
which \name can manage by keeping a track of
trustee status over time.

\paragraph{\bf Share integrity}
\label{sec:disc:integrity}
\name protects the confidentiality of $\key$ but not
the integrity of shares.
By corrupting the trustees of a single subsecret,
a below-threshold adversary can tamper with that subsecret
undetectably; since all subsecrets are needed,
$\key$ cannot be reconstructed.
This attack is akin to denial-of-service,
as the adversary prevents a user from obtaining shares
of the key and thus, this attack could be executed by
simply preventing the trustees from sending shares to the user.
Due to this reason,
adding verifiability~\cite{feldman1987practical} to MLSS does not
address this issue.
We instead rely on redundancy
and introduce thresholded MLSS in \cref{app:thresholded-mlss},
which uses Shamir-shares $\key$ for subsecrets,
and ensures that recovery tolerates corruption
in subsecrets as well.

\section{Implementation and Evaluation}
\label{sec:eval}
In this section,
we answer the following questions:
\begin{itemize}[leftmargin=*]
    \item
        What is the overhead introduced due to
        the use of anonymity set and MLSS in \name
        (\cref{sec:eval:overhead})? 
    \item
        How does \name's recovery time scale with large thresholds
        and large address books
        (\cref{sec:eval:practical})?
    \item
        To what extent does \name reduce an adversary's chances
        of recovering a user's key
        (\cref{sec:eval:security})?
\end{itemize}

\subsection{Implementation}
We implemented \name%
\footnote{The anonymized code is available
\href{https://anonymous.4open.science/r/apollo\_asiaccs\_2027-17CA/}
{here}.}
using Go version 1.22~\cite{meyerson2014go},
in 7655 lines of code
as counted by CLOC~\cite{cloc}.
We use SHA256 as hash function.
The key, $\key$, is an array of elements from
binary extension field of degree $16$.

\subsection{Experimental Configuration}
We ran the experiments on
AMD EPYC 7702 Processor, with 16GB RAM and 4 cores. 

To evaluate the performance gains of the multi-layered approach
(\cref{sec:apollo:mlss})
over the single-layered approach
(\cref{sec:apollo:single}),
we measure CPU and wall-clock recovery times.
CPU time reflects the computational cost of recovery,
while wall-clock time captures the delay between
user reconnection and vault recovery.
Since both follow the same trend, we report only CPU time.

To quantify a user's advantage over an adversary in key recovery,
we simulate the key recovery protocol and compute the recovery probability
as the fraction of successful runs.
We report the cumulative distribution function (CDF) of this probability,
with each value computed from 1 million simulation runs.

\subsection{Overhead of anonymity set}
\label{sec:eval:overhead}
We evaluate the overhead due to chaff data in terms of
(i) computation time after each reconnection
and (ii) bandwidth consumed during backup.
We use the following parameters:
(i) recovery threshold of 50\% ($\tau$),
(ii) an absolute threshold (\cref{sec:apollo:mlss}) of $3$ for MLSS,
(iii) $20$ trustees,
based on a study on social ties between
people~\cite{pewties}.

\begin{figure}[htb]
  \centering
    \includegraphics[width=0.28\textwidth]
    {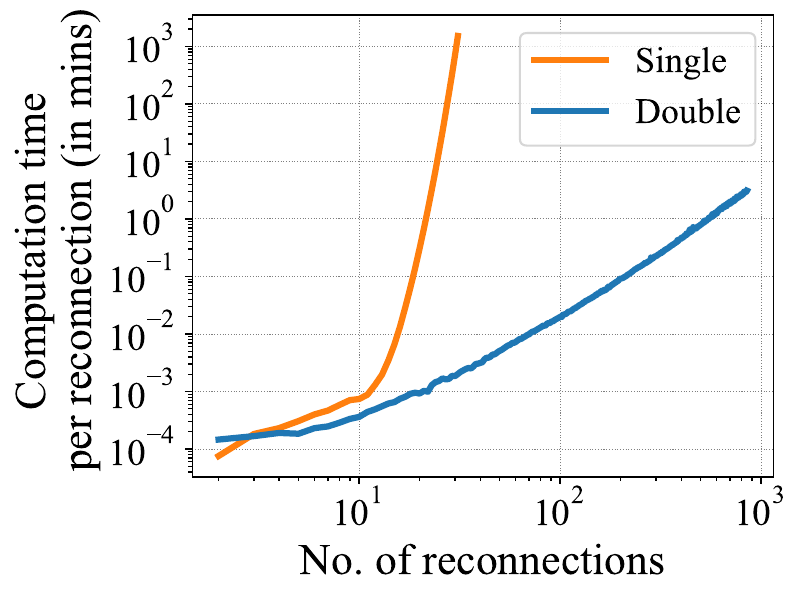}
    \caption{Computation time after a new reconnection;
    the time taken by single-layered approach grows rapidly due to
    the exponential growth in the no. of combinations,
    requiring more than $25$ hours after $30$ reconnections,
    while the double-layered approach needs only $155$ seconds
    after $850$.}
    \label{fig:eval:reconnect}
\end{figure}

\cref{fig:eval:reconnect} presents the
time taken by the MLSS recovery algorithm to compute over
all share combinations after a new reconnection,
in comparison with the single-layered approach.
At the $30$-th reconnection,
MLSS shows an improvement of $5$ orders of magnitude.
However, for the first 3 reconnections,
the computation time of the single-layered approach
is $0.7\times-0.9\times$
of that using MLSS.
This is because each contact holds two shares in our implementation
of MLSS, compared to one in the single-layered approach,
resulting in more computation for \multiss in the first few
reconnections.
Beyond the $4$th reconnection,
the impact of exponential growth in the number of combinations
in the single-layered approach
dominates over
the impact of multiple shares held by each contact in MLSS,
leading to a $\approx 740k\times$ improvement at $30$ reconnections.

\begin{figure}[htb]
    \centering
        \includegraphics[width=0.25\textwidth]
        {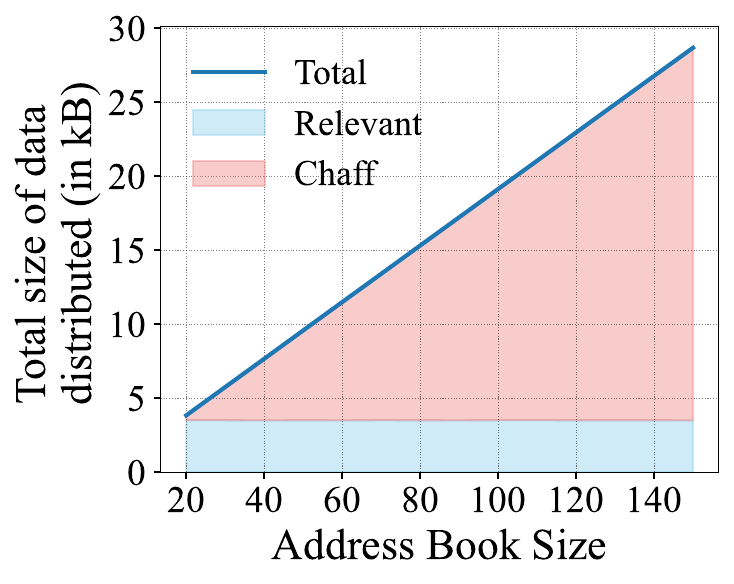}
        \caption{Size of data distributed across the address book.}
        \label{fig:eval:packet}
\end{figure}

\cref{fig:eval:packet} depicts the variation of
the total bandwidth consumed
during key backup.
The bandwidth grows linearly with the address book size due to the
increasing amount of chaff data.
However, it remains below $30$kB even for $150$ contacts,
since MLSS limits each contact to two shares
(\cref{sec:apollo:mlss:num-ss}).

\subsection{Ablation Study}
\label{sec:eval:practical}
We evaluate MLSS recovery algorithm's scalability
using the expected cumulative recovery time:
the total time spent computing share combinations
until the key is recovered,
averaged over $100$ runs.
Through these experiments,
we study the impact of address book size,
percentage threshold, and absolute threshold on recovery time.
We use the same parameters as \cref{sec:eval:overhead}
and an address book of size $150$
(Dunbar’s number~\cite{dunbar2010many}).
However, for the single-layered
approach and MLSS with absolute threshold $> 3$,
we use address book size $<100$
since some runs took more than $10$ hours.

\begin{figure}[htb]
    \centering
        \subfloat[Address book size]{\includegraphics[width=0.242\textwidth]
        {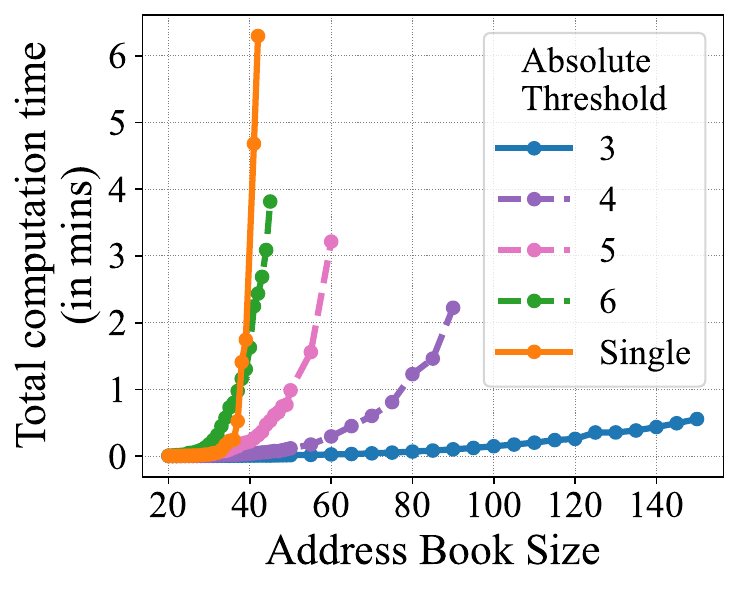}%
        \label{fig:eval:at}}
        \hfil
        \subfloat[Percentage Threshold of the key]
        {\includegraphics[width=0.233\textwidth]
        {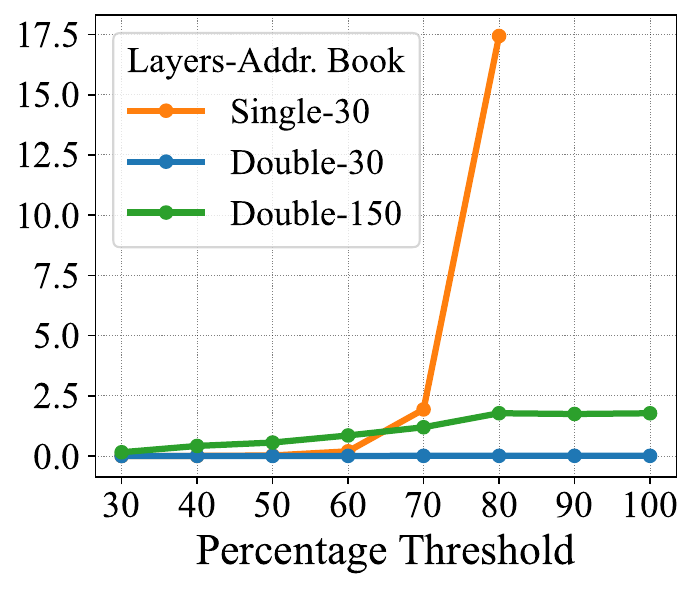}%
        \label{fig:eval:threshold}}
        \caption{
            Total computation time:
            (a) the time taken with single layer grows
            exponentially due to the growth in no. of
            combinations.
            With double layer,
            polynomial growth in the no. of combinations
            leads to rapid growth in recovery time
            for thresholds $>3$.
            (b) the time taken with single layer grows
            exponentially,
            while the double-layered approach remains almost unaffected
            (even for $5\times$ address book size)
            because a small threshold ($3$) is used for
            generating share combinations.
        }
        \label{fig:eval:param}
\end{figure}

\cref{fig:eval:at} shows the impact of using
larger absolute thresholds on
MLSS's cumulative recovery time,
in comparison with the single-layered approach.
If a user has $42$ contacts,
MLSS (absolute threshold $3$) is expected to take $0.59$s,
while the single-layered approach is expected to run for $378$s,
over all reconnections.
The computation time increases rapidly
for absolute thresholds $>3$,
justifying our choice of using an absolute threshold
of $3$.
\cref{fig:eval:threshold} shows
the variation of cumulative recovery time
with the threshold.
Even while using a
$5\times$ larger ($150$) address book,
MLSS takes $7\times$ less time for recovery for threshold $\geq 80$.
This reduction is due to
the use of a small absolute threshold ($3$) in MLSS,
which eliminates the exponential growth in
share combinations generation.

\subsection{Security against malicious recovery}
\label{sec:eval:security}
We use the same parameters as \cref{sec:eval:overhead}
and an address book of size $150$.
We analyze \name's defense against both
the baseline $f$-of-$n$ adversary (\cref{sec:overview:threat}),
and realistic adversaries (\cref{sec:disc:realistic}).

\begin{figure}[htb]
    \centering
        \subfloat[Address book]{\includegraphics[width=0.248\textwidth]
        {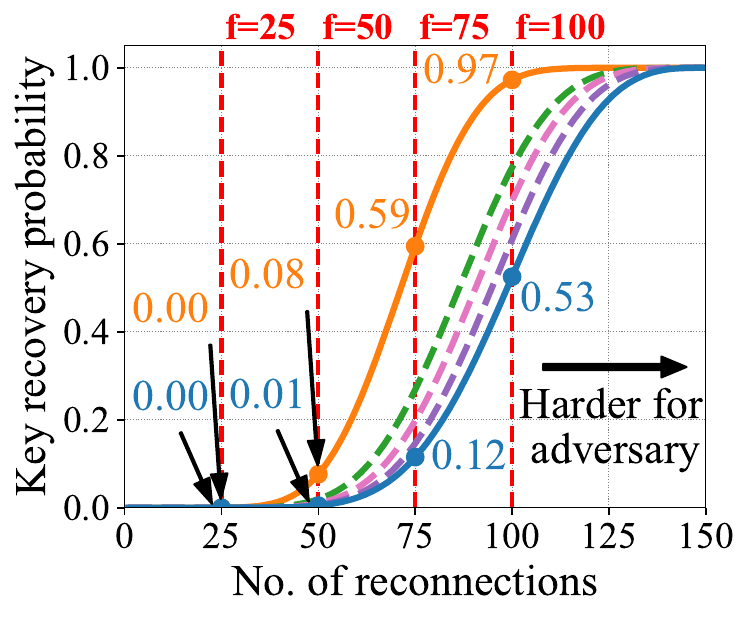}%
        \label{fig:eval:sec:baseline:anonymity}}
        \hfil
        \subfloat[Trustees]
        {\includegraphics[width=0.228\textwidth]
        {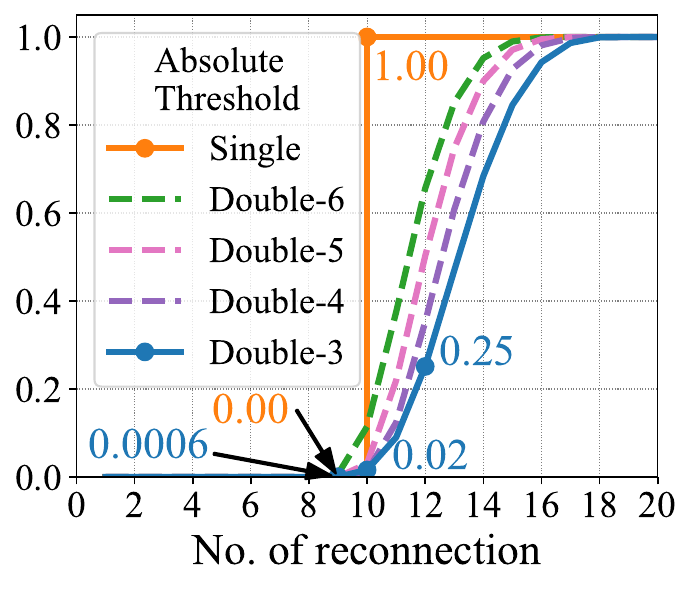}%
        \label{fig:eval:sec:baseline:trustees}}
        \caption{
            Security in baseline model:
            the solid lines in (a) show adversary
            needs to put in considerable effort
            to have a non-zero chance for recovery ($f>25$);
            the dashed lines in (a) and (b) show
            that increasing the absolute threshold for each subsecret
            brings \multiss's design closer to the single-layered approach.
            Although (b) shows a $0.06\%$
            recovery probability after reconnecting with
            $9$ ($<50\% \text{ of } 20$) trustees,
            both (a) and (b) show that \multiss typically reduces recovery
            probability compared to a single-layer design,
            thereby increasing the effort required
            from both the user and the adversary
        }
        \label{fig:eval:security:baseline}
\end{figure}

\paragraph{\bf Security against baseline adversary.}
\cref{fig:eval:security:baseline}
depicts \name's security in the baseline threat model
(\cref{sec:overview:threat}).
The recovery probability in the double-layered design
deviates from the single-layered design due to the
probabilistic recovery (\cref{sec:apollo:mlss:prob}):
we observe $0.05\%$ probability of recovery after
reconnecting with $9 (< 50\% \text{ of } 20)$ trustees;
however, due to the same reason,
the probability of \multiss stays
below that of single-layered design for more trustees.
Similarly, for a given number of reconnections,
the probability of \multiss is always less than the
single-layered approach---demanding more effort from
both the user and the adversary.
Additionally, increasing the absolute threshold for each subsecret
brings the distribution of \multiss closer to the single-layered approach.
Finally, \name makes it hard for an adversary to recover
as she needs to continue compromising contacts
without knowing the trustees;
however, the baseline model does not quantify this security gain.

\paragraph{\bf Security against realistic adversary}
\label{sec:eval:security:realistic}
We assume
a user remembers $40\%$ of her trustees---%
based on Ebbinghaus' curve~\cite{murre2015replication}.
While trying to acquire shares,
the adversary faces resistance due to the following
factors highlighted above (\cref{sec:disc:realistic}):
(i) uncertainty in obtaining shares;
(ii) error in identifying trustees;
and (iii) risk of getting reported.
We consider three types of adversaries:
(i) moderate adversary: knows $30\%$ trustees
and has $25\%$ chance of obtaining shares,
which is
consistent with a large-scale phishing study~\cite{lain2022phishing}.
We set the probability of an adversary getting reported to $5\%$
which is half the reporting probability in~\cite{lain2022phishing};
(ii) invasive adversary: same as the moderate adversary but knows
as many trustees as the user ($40\%$);
(iii) coercive adversary: same as the moderate adversary but
has a $50\%$ extortion rate.
Our analysis also
includes \emph{impunities},
adversaries that never get caught,
for demonstrating
the impact of reporting.
We believe that these percentages
cover most kinds of real-world adversaries;
however,
validating these percentages would require a user study,
which we leave for future work.

\begin{figure}
    \centering
        \includegraphics[width=0.3\textwidth]
        {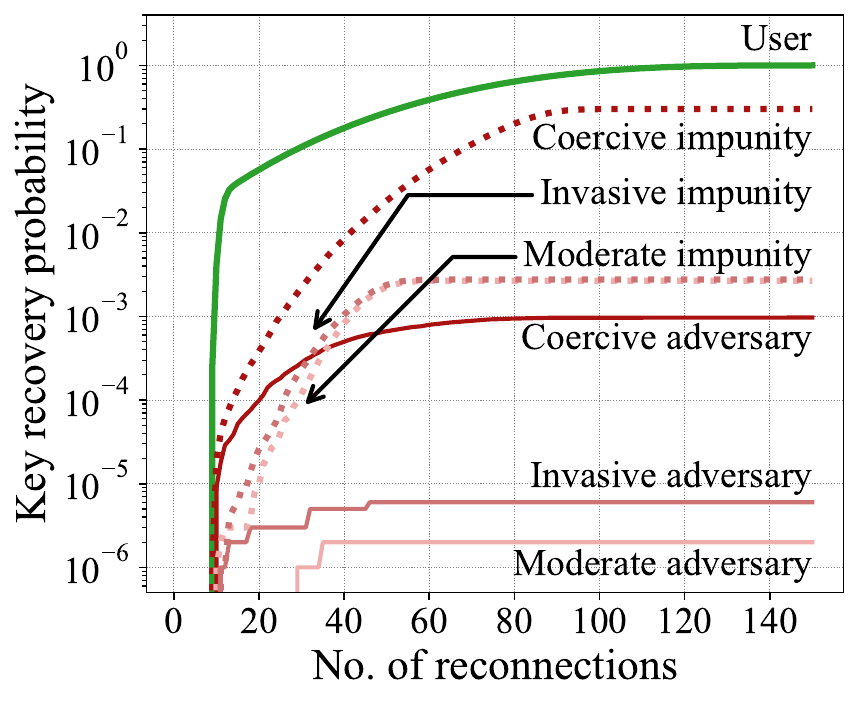}%
        \caption{
            Security against realistic adversaries:
            user remembers $40\%$ of trustees and recovers the key
            as long as she can reconnect;
            a moderate adversary
            knows $30\%$ of trustees, has $25\%$ chance of
            obtaining shares and
            $5\%$ chance of getting reported by a contact,
            while the invasive adversary knows $40\%$ of the trustees and
            the coercive adversary has $50\%$ chance of obtaining shares.
            The corresponding \emph{impunities} depict ideal adversaries that
            never get reported.
        }
        \label{fig:eval:security:obtain}
\end{figure}

\cref{fig:eval:security:obtain} presents \name's
defense against types of adversaries.
For the moderate and invasive adversaries,
the adversary's probability of recovery remains
below $0.001\%$.
For the coercive adversary,
the probability remains below $0.1\%$.
Furthermore, a small reporting probability ($1$-in-$20$ in our case)
reduces adversary's success chances greatly:
while the coercive impunity has $30\%$ chance of
recovering the key,
a real-world adversary has a maximum probability of $0.1\%$.
Hence, \name keeps an adversary's chances of recovering
the key below $0.1\%$,
even for coercive adversaries.

\begin{figure}[htbp]
    \centering
        \includegraphics[width=0.38\textwidth]
        {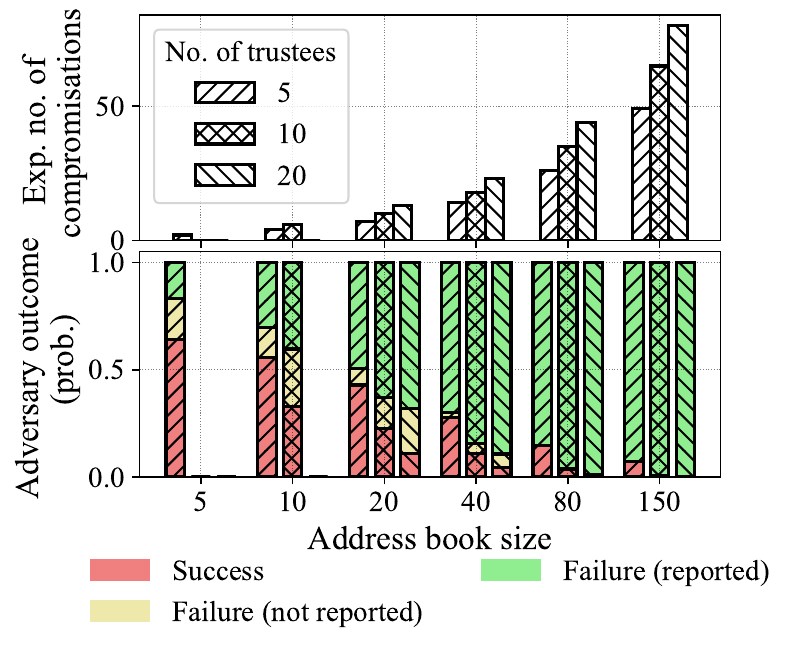}%
        \caption{
            Impact of address book and trustees set size on security
            against a coercive adversary
            ($50\%$ chances of obtaining shares):
            the expected number of people to be compromised
            increases with address book size,
            forcing the adversary to put more effort for recovery;
            as the address book size increases,
            the adversary gets reported often (green bars).
            For existing solutions with \emph{3-out-of-5} configuration,
            the adversary might succeed with $>7\%$
            even for address book size 150.
        }
        \label{fig:eval:security:combined}
\end{figure}

\cref{fig:eval:security:combined} depicts the
impact of the address book and trustee set size
on security.
While using the $3$-of-$5$ configuration of existing approaches,
there is $>7\%$ chances of adversarial recovery,
even for address book size $150$.
On the other hand,
for trustee set sizes $10$ and $20$
along with an address book of size $>80$,
the adversary has only $<0.5\%$ chances of succeeding.
Therefore, while the anonymity set
reduces the adversarial success probability,
it is also important to allow users to choose their trustees
according to their needs since using a
small fixed configuration, like $3$-of-$5$ in existing approaches,
does not ensure protection against a coercive adversary.
Finally, using Dunbar's number~\cite{dunbar2010many}
in our security analysis describes a scenario
convenient for the adversary, not for \name:
as \cref{fig:eval:security:combined} depicts,
\name's security guarantees only improve
with more trustees or contacts.

Our experiments show \name:
(i) can recover a user's key within three minutes of
reconnecting with a new contact;
(ii) ensures key recovery as long as the user
continues to reconnect;
and (iii) diminishes the chances of recovery $< 0.1\%$,
even for strong adversaries.

\section{Related Work}\label{sec:related}

\paragraph{\bf Secret backup}
CanDID~\cite{maram2021candid} enables key recovery
via a secret management committee,
based on login proofs from a threshold of platforms.
This creates a circular dependency---using one secret
for another.
ANARKey~\cite{kate2025anarkey},
Acsesor~\cite{chase2022acsesor} and CALYPSO~\cite{kokoris2020calypso}
enable decentralized key management
but rely on a blockchain.
\name neither requires a user to remember another secret
nor does it rely on a blockchain.
Adei et al.~\cite{adei2023how} deal with
vault recovery in the event of device loss,
but require users to monitor
for illegitimate attempts
to prevent malicious recovery.
\name does not require the active involvement of the user
for security.
There are solutions that
back up keys with
multiple trusted
hardware devices~\cite{dauterman2020safetypin,connell2024secret},
however they are vulnerable to security risks of trusted hardware%
~\cite{jauernig2020trusted,nilsson2020survey}.

\paragraph{\bf Secret sharing variants}
\name does not need verifiable secret sharing~\cite{feldman1987practical}
since the user (dealer) needs to distribute correct shares.
Password-protected secret
sharing~\cite{bagherzandi2011password,jarecki2016highly}
uses a password to
ensure security even
if all trustees are compromised,
but it requires a user to remember a password,
introducing a circular problem.
Hierarchical secret sharing schemes%
~\cite{traverso2016dynamic,tassa2007hierarchical,
farras2012ideal} employ
multiple layers of shares
to create a hierarchy of trustees;
nested Secret Sharing~\cite{galletta2021overcoming}
uses layering to amplify computation for preventing
adversarial activities.
In contrast, \name uses MLSS only for scalability.
Proactive secret sharing~\cite{herzberg1995proactive,maram2019churp,
yurek2023long} enables a set of parties to
update the secret shares.
\name does not need PSS
since the share update has to be handled by the user
(not by the contacts).
A recent work, Chorus~\cite{rathee2026chorus},
offers strong privacy guarantees for secret recovery,
but lacks user control as the committee is chosen
by the protocol, not by the user.
Eldridge et al.~\cite{eldridge2024abuse} introduce MDSS,
which reconstructs secrets in the presence of relevant and irrelevant
indistinguishable shares.
MDSS and MLSS serve orthogonal goals:
the former ensures unlinkability of shares
across users and uses error-correcting code,
while the latter solely focuses on reducing combinatorial overhead
by using multiple layers.

\paragraph{\bf Social authentication}
Some applications
use security questions
for  account recovery~\cite{hang2015have, hang2015know},
but pose serious usability and security
issues~\cite{bonneau2015secrets, rabkin2008personal}.
Facebook introduced a mechanism called
social authentication~\cite{gong2014security},
where a user
must identify known people
in a set of photos to regain access.
However, this method
is vulnerable to attacks
using social engineering and artificial
intelligence~\cite{alomar2017social, polakis2012all, kim2012social}.
In contrast,
\name increases the potential targets for an adversary,
making such attacks harder to carry out.

\section{Conclusion}\label{sec:conclusion}
This paper introduces the notion of recovery metadata in social recovery,
emphasizing the importance of concurrently ensuring self-recovery and
recovery metadata privacy.
To achieve the illustrated two-fold goal,
we proposed
\name, the first social key recovery protocol
that minimizes the memorability burden,
while ensuring metadata privacy.
Our security analysis and performance evaluation confirm that
\name ensures vault recovery for users and
offers strong protection against malicious recovery attempts,
while maintaining low latency during recovery.

\begin{acks}
We thank Adway Girish for helpful conversation and feedback
on bounds for probability.
We thank Shefali Gupta for inputs on psychology
and memory research.
We thank Albert Kwon,
Bernhard Tellenbach and Saiid El Hajj Chehade for their
helpful feedback on early drafts of this paper.
This research is supported by armasuisse Science and Technology and
the AXA Research Fund.
The authors used ChatGPT and Claude
in all the sections to revise the writing style.
In particular, they used it to reduce the verbosity of sentences
and finding alternatives for improving the presentation.
\end{acks}

\bibliographystyle{ACM-Reference-Format}
\bibliography{references}

\appendix
\crefalias{section}{appsec}
\crefalias{subsection}{appsec}

\section{Open Science}
\label{app:open-science}
Our code and simulation outputs are available
\href{https://anonymous.4open.science/r/apollo\_asiaccs\_2027-17CA/}
{\color{blue}{here}}.

\section{Analysis of Single-Layered Approach}
\label{app:security:single}
\begin{figure}[htb]
    \centering
    \subfloat[Single (\cref{sec:apollo:single})]
        {\includegraphics[width=0.242\textwidth]
        {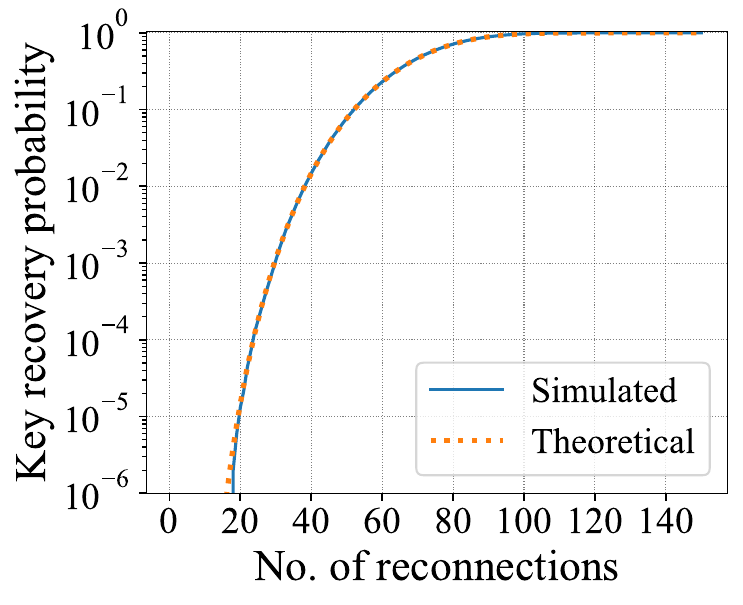}%
        \label{fig:eval:sim-th:single}}
        \hfil
        \subfloat[MLSS (\cref{sec:apollo:mlss})]
        {\includegraphics[width=0.233\textwidth]
        {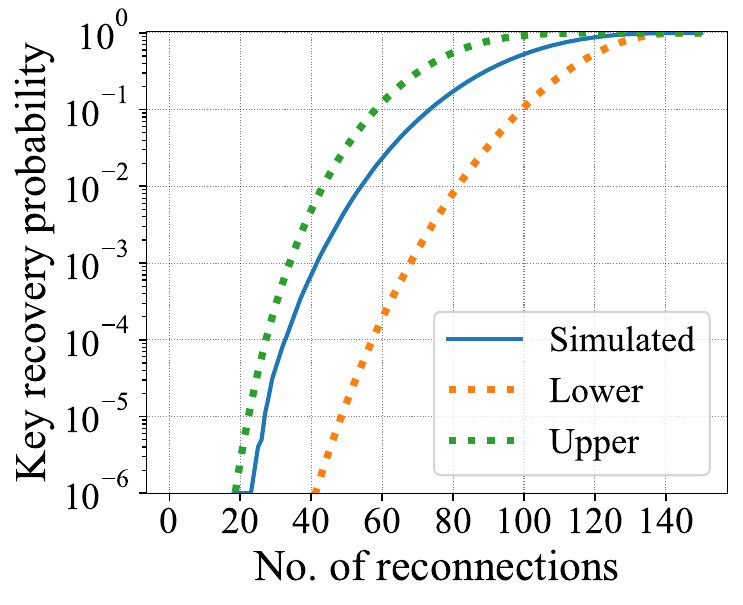}%
        \label{fig:eval:sim-th:double}}
        \caption{Theoretical and empirical analysis comparison.}
        \label{fig:eval:sim-th}
\end{figure}

Here,
we provide a theoretical analysis of
the recovery probability
single-layered design (\cref{sec:apollo:single}).
We use
the notations provided in~\cref{tab:notation-app}.
In our analysis,
we assume that the contacts are approached in an arbitrary order, i.e.,
the next contact which is to be approached is
chosen in a uniformly random manner
without replacement.

\begin{table}[!t]
    \centering
    \caption{Notations}
    \label{tab:notation-app}
    \begin{tabular}{c p{7cm}}
      \toprule
      \textbf{Notation} & \textbf{Description} \\
      \midrule
      $\user$ &
      User \\
      $\adversary$ &
      Adversary \\
      $\alpha$ &
      Absolute threshold \\
      $\beta$ &
      No. of subsecrets \\
      $\gamma$ &
      No. of shares held per person ($\gamma$ is an integer $\geq 2$) \\
      $\delta$ &
      Fraction of people with $\gamma$ key shares\\
      $\tau$ &
      Fraction of trustees for recovery\\
      $\numtrusteesset$ &
      No. of trustees \\
      $\numaddrbook$ &
      No. of contacts \\
      $k$ &
      No. of people contacted \\
      $\numtrusteessetrec$ &
      Minimum no. of trustees to be contacted for recovery\\
      $\numtrusteessetreconly$ &
      No. of trustees among $\numtrusteessetrec$ trustees that hold $\gamma$
      key shares \\
      \bottomrule
    \end{tabular}
\end{table}

\subsection{Key recovery probability}
The probability of recovering the
key at the $k$-th reconnection is given by:

\begin{align}
    \mathrm{P}(X = k) &= 
                      t \times (k-1)! \times 
    \frac{{\numtrusteesset \choose t}%
    {{\numaddrbook - \numtrusteesset} \choose {k - t}}}%
    {n \times (n-1) \times \dots (n-k+1)} \nonumber \\
                      &= \frac{t}{k} \times%
    \frac{{\numtrusteesset \choose t}%
    {{\numaddrbook - \numtrusteesset} \choose {k - t}}}%
    {{n \choose k}}
    \label{eqn:single-layered-exact}
\end{align}

\cref{eqn:single-layered-exact} gives us
the probability of recovering the key
at the $k$-th person,
given the contacts are approached randomly.
To validate our analysis,
we compare our analysis with the simulated
results in \cref{fig:eval:sim-th:single}
and as depicted,
our analysis is accurate,
matching with
the simulation exactly.

\section{Analysis of Multi-Layered Secret Sharing}
\label{app:security:multi}
\subsection{Bounds on key recovery probability}
Here,
similar to the single-layered analysis (\cref{app:security:single}),
we consider contacts are approached in
an arbitrary order.
For the multi-layered setting (MLSS),
the probability evaluation needs to consider:
(i) the number of trustees contacted and,
(ii) based on the number of trustees contacted,
we need to evaluate the probability of key recovery.
Hence, the probability of key recovery can be segregated into two parts:

\begin{itemize}[leftmargin=*]
    \item $\mathrm{P}_1(X=k, Y=k_T)$ = Probability of contacting $k_T$ trustees
        after contacting $k$ contacts.

    \item $\mathrm{P}_2(Y=k_T)$ = Probability of recovering the key
        at the $k_T$-th trustee.
\end{itemize}

The probability of recovering at the $k$-th contact
is given by:
\begin{align}
    \mathrm{P}(X=k) = \sum_{k_T}\mathrm{P}_1 \times \mathrm{P}_2
    \label{eqn:double-layered}
\end{align}

The exact value of $P_1$ is obtained using~\cref{eqn:single-layered-exact}:

\begin{align}
    \mathrm{P}_1(X=k) = \frac{k_{T}}{k} \times
    \frac{{\numtrusteesset \choose k_{T}}%
    {{\numaddrbook - \numtrusteesset} \choose {k - k_{T}}}}%
    {{\numaddrbook \choose k}}
\end{align}

In contrast,
obtaining a closed-form equation for $\mathrm{P}_2$
is intractable since
counting each of the successful cases is hard.
Intuitively, this counting is hard as we need to
treat each share of a subsecret as a distinct share,
due to which a generalized equation for counting
leads to many repetitions.
Therefore, for $\mathrm{P}_2$,
we provide bounds instead and
compare the bounds with our simulated results.

\paragraph{\bf Lower bound for $\mathrm{P}_2$}
The lower for $\mathrm{P}_2$ is given by:

\begin{align}
    \mathrm{P}_{2,\text{lower}}(X \leq k_T) =%
    \frac{
        {{\lfloor \frac{\absolute}{\threshold} \rfloor}%
        \choose \absolute}%
        ^{\numss} \times%
        {{\numtrusteesset \spp - \absolute \numss}%
        \choose {k_T \spp - \absolute \numss}}
    }
    {
        {\max_{i}{\left({\absolute \choose i}%
        {{\lfloor \frac{\absolute}{\threshold} \rfloor - \absolute}%
        \choose i}\right)}}^\numss%
        \times {{\numtrusteesset \spp} \choose {k_T \spp}}
    }
    \label{eqn:double:lower}
\end{align}

We obtain $\mathrm{P}_{2,\text{lower}}(X = k_T) =
\mathrm{P}_{2,\text{lower}}(X \leq k_T) -
\mathrm{P}_{2,\text{lower}}(X \leq (k_T - 1))$.
The value obtained from this is essentially the lower bound
because we are dividing by the maximum possible number of repititions
of a combination for each subsecret.

\paragraph{\bf Upper bound for $P_2$}
For the upper bound,
we remove the $\max_i$ term, which gives:

\begin{align}
    \mathrm{P}_{2,\text{upper}}(X \leq k_T) =%
    \frac{
        {{\lfloor \frac{\absolute}{\threshold} \rfloor}%
        \choose \absolute}%
        ^{\numss} \times%
        {{\numtrusteesset \spp - \absolute \numss}%
        \choose {k_T \spp - \absolute \numss}}
    }
    {
        {{\numtrusteesset \spp} \choose {k_T \spp}}
    }
    \label{eqn:double:upper}
\end{align}

We plugin the values
obtained from \cref{eqn:double:lower,eqn:double:upper}
into \cref{eqn:double-layered}
and obtain the lower bound and upper bound.
We show the bounds for the two-layered setting
in \cref{fig:eval:sim-th:double}.

Next, we analyze the security of the multi-layered setting with respect to the
threshold set by the user (${\threshold}$).

\begin{table}[!t]
    \centering
    \caption{Default Values}
    \label{tab:default-values}
    \begin{tabular}{p{5.5cm} c}
      \toprule
      \textbf{Parameter} & \textbf{Default Value} \\
      \midrule
      No. of trustees & 20 \\
      Anonymity Set Size & 150 \\
      Percentage threshold ($\tau$) & 0.5 \\
      Absolute Threshold ($\alpha$) & 3 \\
      No. of shares per person ($\gamma$) & 2 \\
      No. of subsecrets ($\beta$) & 6 \\
      \bottomrule
    \end{tabular}
\end{table}

\subsection{Choosing the number of shares per subsecret}
We assign $\lfloor \frac{{\absolute}}{{\threshold}} \rfloor$ shares to each
subsecret. 
The following observation is useful:
\begin{align}
    \floor*{\frac{{\absolute}}{{\threshold}}} \leq 
    \frac{{\absolute}}{{\threshold}} 
    \implies \frac{{\absolute}}{\lfloor \frac{{\absolute}}{{\threshold}} 
    \rfloor} \geq {\threshold}.
    \label{eq:comp-floor}
\end{align}

\cref{eq:comp-floor} implies that choosing $\lfloor . \rfloor$ guarantees that
the threshold of each subsecret is at least ${\threshold}$.
This guarantee would not have been provided with
$\lfloor . \rceil$ or $\lceil . \rceil$.
If everyone held one share, then this design would have provided
\emph{perfect security}, i.e.,
the same security guarantees as the single-layered approach.
Since it is not the case in \multiss, we analyze the
implications of share distribution in \multiss.

\subsection{Perfect Security Guarantees}
\label{app:security:multi:perfect}
Given the
distribution in \multiss, a trustee either receives $\spp$ or $(\spp - 1)$ key
shares. Therefore, we have:
\begin{align}
    \spp \numtrusteessetonly +
    (\spp - 1) (\numtrusteesset - \numtrusteessetonly)
    = \floor*{\frac{{\absolute}}{{\threshold}}} \beta,
    \label{eq:shares-sum}
\end{align}
which gives us:

\begin{align}
    \beta =
    \frac{\floor*{ \numtrusteesset(\delta + \spp - 1) }}
    {\floor*{ \frac{{\absolute}}{{\threshold}} }}.
    \label{eq:beta-val}
\end{align}

For recovering the key $\key$, $\user$ needs to at least obtain ${\absolute}$
number of shares of each subsecret:
\begin{align}
    \spp \numtrusteessetreconly + 
    (\spp - 1)(\numtrusteessetrec - \numtrusteessetreconly) \geq
    {\absolute} \beta,
    \label{eq:recovery-inequality}
\end{align}

For evaluating the bounds,
we have two cases
from~\cref{eq:recovery-inequality}: (i) $\spp \numtrusteessetonly \geq
{\absolute} \beta$; and (ii) $\spp \numtrusteessetonly < {\absolute} \beta$.

\paragraph{\bf Case 1, $\enspace \spp \numtrusteessetonly \geq {\absolute} 
\beta$} In this case, $\numtrusteessetrec = \numtrusteessetreconly \leq
\numtrusteessetonly$. Therefore, from~\cref{eq:recovery-inequality}, we have
$\spp \numtrusteessetrec \geq {\absolute} \beta$. Using~\cref{eq:beta-val}
and~\cref{eq:comp-floor} respectively, we have 
\begin{align}
	\spp \numtrusteessetrec &\geq \frac{{\absolute}}
	{\floor*{ \frac{{\absolute}}{{\threshold}} }}
	\floor*{ \numtrusteesset(\delta + \spp - 1) }
	\nonumber
\end{align}

, which gives us:

\begin{align}
    \numtrusteessetrec 
     &> \threshold
    \left(\numtrusteesset - \frac{\numtrusteesset(1 - \delta) + 1}{\spp}\right).
    \label{eq:perf-sec-case-1-1}
\end{align}

Next, using~\cref{eq:beta-val} and the relation $x \geq
\floor*{x} > x - 1$,
we have:
\begin{align}
    \spp (\numtrusteessetrec - 1) &\leq \frac{{\absolute}}
    {\lfloor \frac{{\absolute}}{{\threshold}} \rfloor}
    \floor*{ \numtrusteesset(\delta + \spp - 1) } \nonumber
\end{align}

, which gives us:

\begin{align}
   \numtrusteessetrec & < \frac{\threshold}
    {1 - \frac{{\threshold}}{{\absolute}}}
    \left(\numtrusteesset - \frac{\numtrusteesset(1 - \delta)}{\spp}\right)+ 1.
    \label{eq:perf-sec-case-1-2}
\end{align}

\begin{figure}[!t]
    \centering
        \subfloat[$\beta = 6$]
        {\includegraphics[width=0.16\textwidth]
        {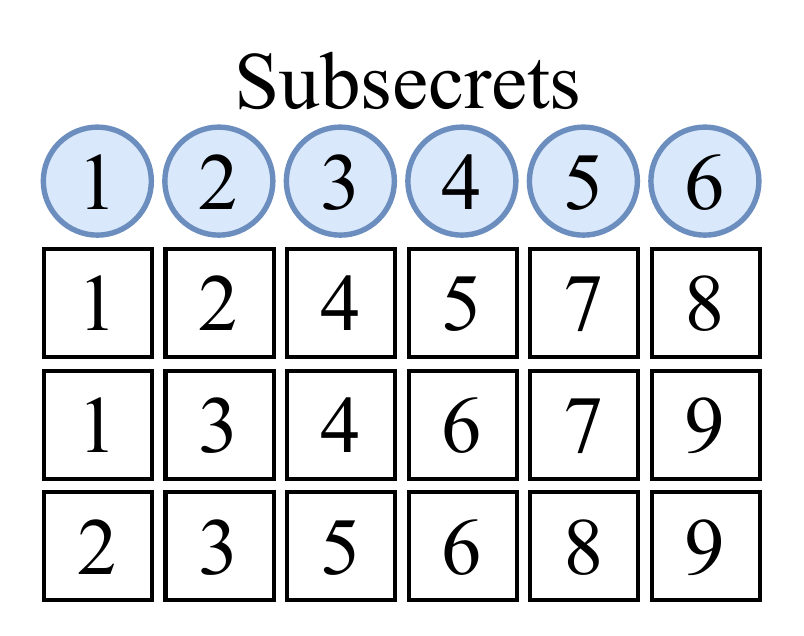}
        \label{fig:app:perf-sec-1-eg-1}}
        \hfill
        \subfloat[$\beta = 5$]
        {\includegraphics[width=0.13\textwidth]
        {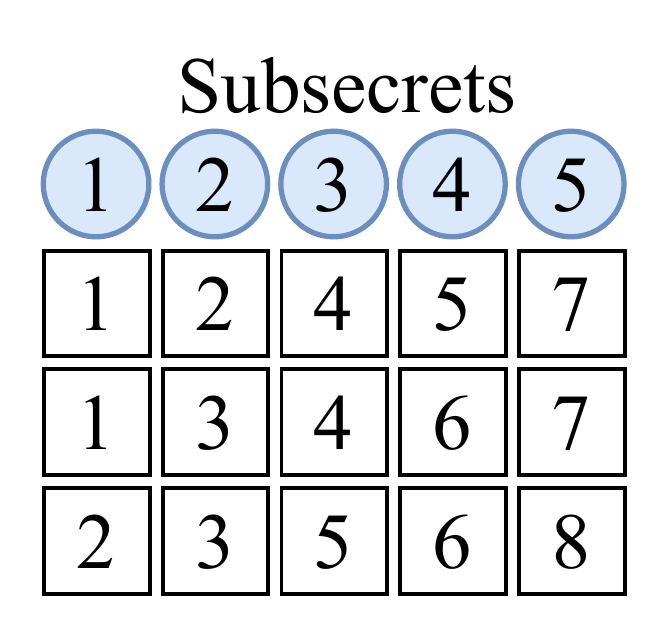}
        \label{fig:app:perf-sec-1-eg-2}}
        \hfill
        \subfloat[$\beta = 4$]
        {\includegraphics[width=0.11\textwidth]
        {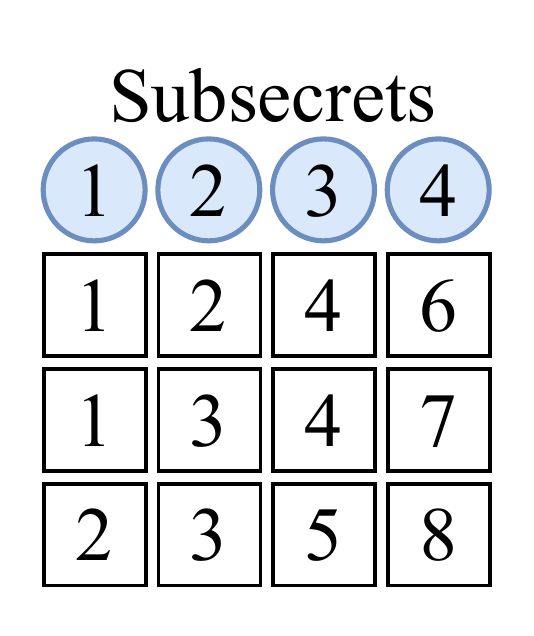}
        \label{fig:app:perf-case-2-eg-2}}
        \caption{Configuration of shares
        obtained to reconstruct the key with less than $\tau \numtrusteesset$
        people.
        The circles depict the subsecrets
        and a tile vertically below the subsecret
        depicts a share of the subsecret.
        The numbers in the tiles depict serial number
        of the trustee from whom the share was obtained.
        }
        \label{fig:eval:perf-sec-1}
\end{figure}

Plugging in the values from \cref{tab:default-values},
we have:
\begin{align}
    \frac{35}{4}
    <
    \numtrusteessetrec
    <
    \frac{59}{5}.
    \label{eq:perf-sec-case-1-eg-1}
\end{align}

We see that this case may not guarantee perfect security because
$\numtrusteessetrec$ can take values lesser than $\tau \numtrusteesset = 10$.
and we show one such configuration in~\cref{fig:app:perf-sec-1-eg-1}.

On the other hand, consider the values of~\cref{tab:default-values}, except with
5 subsecrets, i.e., $\beta = 5$. Then, we have
\begin{align}
    \frac{29}{4}
    <
    \numtrusteessetrec
    <
    10.
    \label{eq:perf-sec-case-1-eg-2}
\end{align} 
We see that this case \emph{cannot} guarantee perfect security
because $\numtrusteessetrec$ is less than $\tau \numtrusteesset = 10$.
and we show one such configuration of recovering
in~\cref{fig:app:perf-sec-1-eg-2}.

As depicted by~\cref{eq:perf-sec-case-1-2},
the lower bound is always less than $\threshold
\numtrusteesset$, but it is closer to the required value if the value of
$\frac{1 - \delta}{\spp}$ is low.
Thus, with a uniform distribution of shares,
the system can achieve
perfect security.
For instance consider the following values: $\numtrusteesset =
20$, $\alpha = 4$, $\beta = 5$, $\gamma = 2$, and $\threshold=0.5$ --- the
number of key shares in this case is $40$ which divides $\numtrusteesset$:
\begin{align}
    \frac{39}{4}
    <
    \numtrusteessetrec
    <
    \frac{87}{7}.
    \label{eq:perf-sec-case-1-eg-3}
\end{align}

\cref{eq:perf-sec-case-1-eg-3} implies that for all the configurations the value
of $\numtrusteessetrec$ is at least $10$, which is $\tau \numtrusteesset$.
Therefore, making the total number of key shares a multiple of the number of
trustees provides the system perfect security.

\paragraph{\bf Case 2, $\enspace \spp \numtrusteessetonly < {\absolute} \beta$} In
this case, we have:
$$  \spp \numtrusteessetonly + (\spp - 1)(\numtrusteessetrec -
\numtrusteessetonly) \geq {\absolute} \beta. $$
, which gives us (from \cref{eq:beta-val,eq:comp-floor}):
\begin{align}
    \numtrusteessetrec &>
    {\threshold}{\numtrusteesset}
    - \frac{\delta \numtrusteesset (1 - \threshold)}{\spp - 1}
    - \frac{\threshold}{\spp - 1},
    \label{eq:perf-sec-case-2-1}
\end{align}

Now, for the upper bound, we have:
$$ \spp \numtrusteessetonly + (\spp - 1)(\numtrusteessetrec - 1 -
\numtrusteessetonly) \leq {\absolute} \beta,$$
which after rearranging and using 
$$(\spp - 1)(\numtrusteessetrec - 1) + \numtrusteessetonly \leq {\absolute}
\beta. $$
Using~\cref{eq:beta-val} to replace $\beta$,
we have:
\begin{align}
    \numtrusteessetrec &<
    \frac{{\threshold}{\numtrusteesset}}{1 - \frac{\threshold}{\alpha}}  
    - \frac{\delta \numtrusteesset \left(1 - \frac{\threshold}{1 - 
    		\frac{\threshold}{\alpha}}\right)}{\spp - 1} 
    + \frac{\spp}{\spp - 1}.
    \label{eq:perf-sec-case-2-2}
\end{align}

As in the previous case, we use these bounds to show examples where perfect
security is not guaranteed. Consider the values used in the simulations
(\cref{tab:default-values}):
\begin{align}
    \frac{15}{2}
    <
    \numtrusteessetrec
    <
    \frac{62}{5}.
    \label{eq:perf-sec-case-2-eg-1}
\end{align}

\cref{fig:app:perf-case-2-eg-2} shows one pattern of obtaining shares such
that $\key$ is reconstructed after contacting less than $\tau \numtrusteesset =
10$ people.
As depicted by~\cref{eq:perf-sec-case-2-2}, $\numtrusteessetrec$ can
take values less than $\tau \numtrusteesset$.
One way of bringing the system
closer to $\tau \numtrusteesset$ is by setting $\delta$ to 0,
which again implies a uniform distribution.

Finally, we highlight that the probability of
perfect security being violated in the two-layered setting is negligible.
This claim is supported by~\cref{fig:app:eval:prob:variation},
where we had no simulation violating the perfect security.

\subsection{Uniform Share Distribution}
\label{app:security:uniform}
\cref{app:security:multi:perfect}
depicts why uniform distribution is useful.
However, they do not justify our choice for
starting with a uniform distribution, i.e.,
\trustees have either $(\spp - 1)$ or $\spp$ shares.
Therefore, now we describe a more skewed version of share
distribution, such that a \trustee can hold $i$ key shares, where $i \in \{1, 2,
\dots, \spp\}$. Let $\delta_i$ be the fraction of \trustees that hold $i$ key
shares, i.e., $\floor*{\delta_i \numtrusteesset}$ hold $i$ key shares.
In this case, we have
\begin{align}
    \sum_{i=1}^{\spp}\floor*{ \delta_i \numtrusteesset } i &=
    \floor*{ \frac{\alpha}{\tau} } \beta, \text{ and} 
    \sum_{i=1}^{\spp}\floor*{ \delta_i \numtrusteesset } &=
    \numtrusteesset.
    \label{eq:trustees-sum:non-uniform}
\end{align}

$$i \left(\numtrusteessetrec - \sum_{j=i+1}^{\spp}\floor*{ \delta_j
\numtrusteesset }\right) + \sum_{j=i+1}^{\spp} j \floor*{ \delta_j
\numtrusteesset }
\geq
\alpha \beta,$$

On solving, we get:
\begin{align}
    \threshold \numtrusteesset >
    \tau \sum_{j=1}^{i-1}\floor*{\delta_j \numtrusteesset} \tau_j
    &+ \tau \floor*{\delta_i \numtrusteesset} \nonumber \\
    + \sum_{j=i+1}^{\spp} \floor*{\delta_j \numtrusteesset}
    ((\tau - 1)(\tau_j) + 1),
    \text{where} r_j = \frac{j}{i}
    \label{eq:shares-sum:inequality-3}
\end{align}

Therefore, $\numtrusteessetrec$ can take values less than $\threshold
\numtrusteesset$.
A way for ensuring $\numtrusteessetrec$ takes values only
greater than $\threshold \numtrusteesset$
is by setting $\delta_j = 0, \forall j
\in \{1, \dots, \spp\} \setminus \{i\}$ and $\delta_i = 1$,
which implies a uniform distribution.

\section{Game-Based Metadata Privacy}
\label{app:security:formal}
\providecommand{\negl}{\mathsf{negl}}
\providecommand{\corrupt}{{C}}
\providecommand{\cfg}{\mathsf{cfg}}
\providecommand{\Advantage}{\mathsf{Adv}}
\providecommand{\qH}{q_{\mathsf{H}}}
\providecommand{\keyspace}{\mathcal{K}}
\providecommand{\Blob}{{B}}

We now make \name's metadata-privacy claim (\cref{sec:problem} and
\cref{sec:discussion}) precise. We model $\mathsf{H}$ as a random oracle with
output length $\lambda$; salts and chaff strings are also $\lambda$-bit, so
real and chaff blobs have identical shape. We consider a \emph{static}
adversary, matching the baseline model of \cref{sec:overview:threat};
integrity under active tampering is out of scope (\cref{sec:discussion}).

Recall (\cref{fig:blob-mlss}) that a trustee holding a share $\share{l,m}$ of
subsecret $\ssmath{l}$ stores the \emph{key-share} blob
$(\share{l,m},\ \salt,\ \Hofstuff{\salt}{\ssmath{l}},\
\Hofstuff{\salt}{\key})$ with a fresh $\salt \randomsample
\{0,1\}^{\lambda}$ per blob, while a \nontrustee\ stores a \emph{chaff} blob
$(\sharesnum, \salt, r_1, r_2)$ with $\sharesnum \randomsample \fieldspace$
and $r_1, r_2 \randomsample \{0,1\}^{\lambda}$. Every contact holds exactly
$\spp$ blobs (\cref{sec:apollo:mlss}), and $\spp < \absolute$
(\cref{tab:default-values}).

\subsection{Metadata privacy}
A \emph{configuration} comprises the recovery metadata: $\cfg =
(\numaddrbook, \numtrusteesset, \threshold)$, or the symbol $\bot$ denoting
a user with no vault, for whom every contact stores $\spp$ chaff blobs. In
the experiment of \cref{fig:game-cfg}, the adversary picks \emph{two}
configurations and a contact, and must tell from the contact's blobs which
configuration the challenger used. Since $\bot$ is a valid configuration,
vault-existence privacy is an instance of the same game.

\begin{figure}[t]
\centering
\fbox{\begin{minipage}{\dimexpr\linewidth-2\fboxsep-2\fboxrule\relax}
\footnotesize
\centering$\mathsf{Exp}^{\mathrm{meta}}_{\adversary}(b)$\par\smallskip
\raggedright
\begin{algorithmic}
    \State $(\cfg^0, \cfg^1, i) \gets \adversary$, where
        $\cfg \in \{(\numaddrbook, \numtrusteesset, \threshold)\} \cup
        \{\bot\}$ and contact $i$ is valid in both configurations
    \State $\key \randomsample \keyspace$;\quad
        $\{\blob{}\} \gets \mathsf{Setup}(\key, \cfg^b)$
    \State $\Blob \gets$ the $\spp$ blobs of contact $i$
    \State $b' \gets \adversary(\Blob)$;\quad \Return $b'$
\end{algorithmic}
\end{minipage}}
\caption{Metadata-privacy experiment: the adversary must tell which of the
two configurations it chose, $\cfg^0$ or $\cfg^1$, produced the blobs.}
\label{fig:game-cfg}
\end{figure}

\begin{definition}[Metadata privacy]
\label[definition]{def:metadata-privacy}
\multiss\ is \emph{metadata-private} if for every PPT adversary $\adversary$,
\begin{align*}
    \Advantage^{\mathrm{meta}}_{\multiss}(\adversary) =
    \big| \Pr\big[\mathsf{Exp}^{\mathrm{meta}}_{\adversary}(0) = 1\big]
    - \Pr\big[\mathsf{Exp}^{\mathrm{meta}}_{\adversary}(1) = 1\big] \big|
    \leq \negl(\lambda).
\end{align*}
\end{definition}

To prove that blobs generated under $\cfg^0$ are indistinguishable from
blobs generated under $\cfg^1$, we show that under \emph{any} configuration
the blobs are indistinguishable from chaff --- a distribution that does not
depend on the configuration. The following lemma states this; it also
captures \emph{trustee privacy}, i.e., that a trustee's blob and a
\nontrustee's blob are indistinguishable.

\begin{lemma}[Blob indistinguishability]
\label[lemma]{lem:blob}
In the random oracle model, no adversary issuing $\qH$ random-oracle queries
distinguishes a key-share blob from a chaff blob with advantage greater than
\begin{align}
    \varepsilon = \qH \big( 2^{-\mu_{\mathrm{ss}}} + 2^{-\mu_{\key}} \big),
    \label{eq:meta-bound}
\end{align}
where $\mu_{\mathrm{ss}}$ and $\mu_{\key}$ are the min-entropies of a
subsecret and of $\key$, respectively, given the shares in the adversary's
view.
\end{lemma}

\begin{proof}
We define three distributions of the challenge blob $\Blob$:
\begin{itemize}[leftmargin=*]
    \item $H_0$: a key-share blob
        $(\share{l,m}, \salt, \Hofstuff{\salt}{\ssmath{l}},
        \Hofstuff{\salt}{\key})$;
    \item $H_1$: as $H_0$, with both tags replaced by independent uniform
        $\lambda$-bit strings;
    \item $H_2$: as $H_1$, with $\share{l,m}$ replaced by a uniform element
        of $\fieldspace$ --- exactly a chaff blob.
\end{itemize}

\emph{$H_0 \approx H_1$.} The two are identical unless $\adversary$ queries
$\mathsf{H}$ at $\salt \,\|\, \ssmath{l}$ or at $\salt \,\|\, \key$. Below
the threshold $\absolute$, $\ssmath{l}$ has min-entropy $\mu_{\mathrm{ss}}$
and $\key$ has min-entropy $\mu_{\key}$ given $\adversary$'s view, so each
query succeeds with probability at most
$2^{-\mu_{\mathrm{ss}}} + 2^{-\mu_{\key}}$; a union bound over the $\qH$
queries gives $|\Pr[H_0] - \Pr[H_1]| \leq \varepsilon$.

\emph{$H_1 \equiv H_2$.} A below-threshold set of Shamir shares is uniform
on $\fieldspace$ and independent of the subsecret, so the two distributions
are identical.

The claim follows by the triangle inequality.
\end{proof}

\begin{theorem}
\label[theorem]{thm:metadata-privacy}
In the random oracle model, \multiss\ is metadata-private:
\begin{align*}
    \Advantage^{\mathrm{meta}}_{\multiss}(\adversary)
    \leq 2 \Big( \spp\, \varepsilon + \binom{\spp}{2} 2^{-\lambda} \Big).
\end{align*}
\end{theorem}

\begin{proof}
Fix $b \in \{0, 1\}$. Condition on the $\spp$ salts of contact $i$ being
distinct, which costs $\binom{\spp}{2} 2^{-\lambda}$. Replace contact $i$'s
key-share blobs by chaff blobs one at a time (chaff blobs need no
replacement); since $\spp < \absolute$, contact $i$'s shares stay below threshold
throughout, so each replacement costs at most $\varepsilon$ by
\cref{lem:blob}. The result --- $\spp$ chaff blobs --- is the same
distribution for $b = 0$ and $b = 1$, as it does not depend on $\cfg^b$;
note both configurations hand over the same number of blobs, since \name\
fixes $\spp$ irrespective of the configuration (\cref{sec:apollo:mlss}).
Applying the above to $b = 0$ and $b = 1$ and the triangle inequality
yields the claim.
\end{proof}

Instantiating \cref{def:metadata-privacy} yields all the properties of
\cref{sec:problem}: $\cfg^1 = \bot$ gives existence privacy, configurations
differing in $\threshold$, $\numtrusteesset$, or $\numaddrbook$ give
configuration privacy, and \cref{lem:blob} gives trustee privacy.
\cref{cor:coalition-privacy} extends the guarantee from one contact to a
coalition. The below-threshold condition is necessary: a coalition holding
$\absolute$ shares of every subsecret can run the recovery algorithm itself,
so no indistinguishability can hold. \cref{sec:eval:security} quantifies the
probability of reaching that regime.

\begin{corollary}[Coalition privacy]
\label[corollary]{cor:coalition-privacy}
\cref{thm:metadata-privacy} holds with the blobs of a set $\corrupt$ of $f$
contacts in place of contact $i$, with advantage at most
$2 \big( \numss\, \qH\, 2^{-\mu_{\mathrm{ss}}} + \qH\, 2^{-\mu_{\key}} +
\binom{f\spp}{2} 2^{-\lambda} \big)$, provided $\corrupt$ holds fewer than
$\absolute$ shares of every subsecret under both configurations.
\end{corollary}

\begin{proof}[Proof sketch]
As in \cref{thm:metadata-privacy}, with the hybrid running over the $f\spp$
blobs of $\corrupt$: since $\corrupt$ stays below threshold throughout,
every subsecret and $\key$ retain min-entropy $\mu_{\mathrm{ss}}$ and
$\mu_{\key}$ given all shares in $\corrupt$, and the oracle queries are
counted once against each of the $\numss$ subsecrets and $\key$.
\end{proof}

\subsection{Key confidentiality}
Confidentiality of $\key$ is information-theoretic and holds under a weaker
condition: $\corrupt$ only needs to be below threshold on \emph{one}
subsecret.

\begin{proposition}[Key confidentiality]
\label[proposition]{prop:confidentiality}
If $\corrupt$ holds fewer than $\absolute$ shares of at least one subsecret,
then the shares held by $\corrupt$ are distributed independently of $\key$.
\end{proposition}

\begin{proof}
That subsecret's below-threshold shares are uniform and independent of the
subsecret (Shamir), and $\key$ is the sum of all $\numss$ subsecrets
(additive sharing); a uniform independent summand decouples the view from
$\key$.
\end{proof}

\paragraph{\bf Remark}
The tag $\Hofstuff{\salt}{\key}$ is \emph{key-confirming} by design: it
is how recovery detects success (\cref{sec:mlss:reconstruction}).
Consequently, an adversary holding a candidate for $\key$ can test it with
one oracle query, and full blobs hide $\key$ only computationally, with the
$\qH\, 2^{-\mu_{\key}}$ term of \cref{eq:meta-bound}. This poses no risk for
a uniformly random vault key, but it implies $\key$ must not be low-entropy,
e.g., derived from a password.

\subsection{Design invariants}
The proofs rely on three conditions that \name\ must maintain.
\begin{itemize}[leftmargin=*]
    \item \emph{Fresh per-blob salt.} Reusing a $\salt$ across two shares of
        the same subsecret would make their tags equal and reveal
        co-membership; the birthday terms in \cref{thm:metadata-privacy}
        and \cref{cor:coalition-privacy} account for accidental collisions.
    \item \emph{Uniform blob count.} Every contact must hold exactly $\spp$
        blobs; if \trustees\ held a different number than \nontrustees, the
        count alone would distinguish them, and
        \cref{thm:metadata-privacy} would fail.
    \item \emph{Sufficient subsecret entropy.} \cref{eq:meta-bound} is
        vacuous unless $\mu_{\mathrm{ss}} \geq \lambda$. As $\key$ lies in
        $\galois{2}{16}$, each subsecret must span enough field elements
        that $\mu_{\mathrm{ss}} \geq \lambda$ given any below-threshold
        view.
\end{itemize}
The bounds hold for static corruption; adaptive corruption requires
programming the oracle and is left as future work.

\section{Variation from single-layered}
\label{app:practical:variation}

\begin{figure}[htbp]
    \centering
        \subfloat[]
        {\includegraphics[width=0.22\textwidth]
        {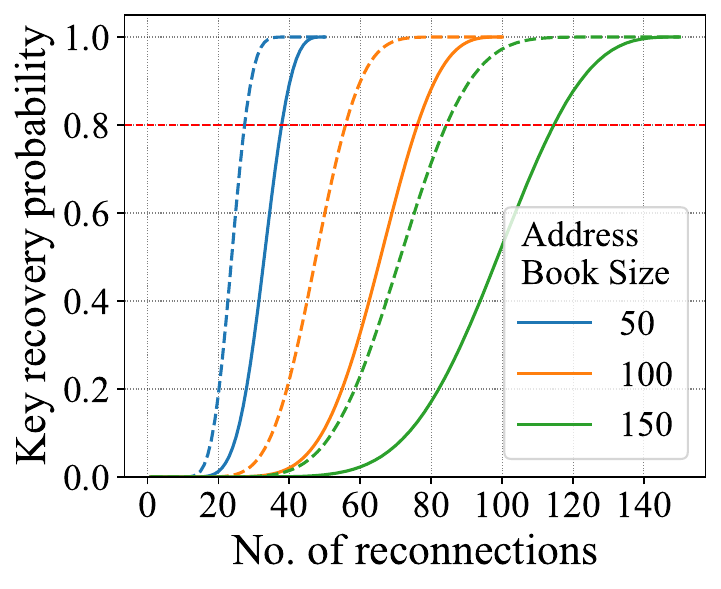}%
        \label{fig:app:eval:prob:variation}}
        \hfil
        \subfloat[]
        {\includegraphics[width=0.22\textwidth]
        {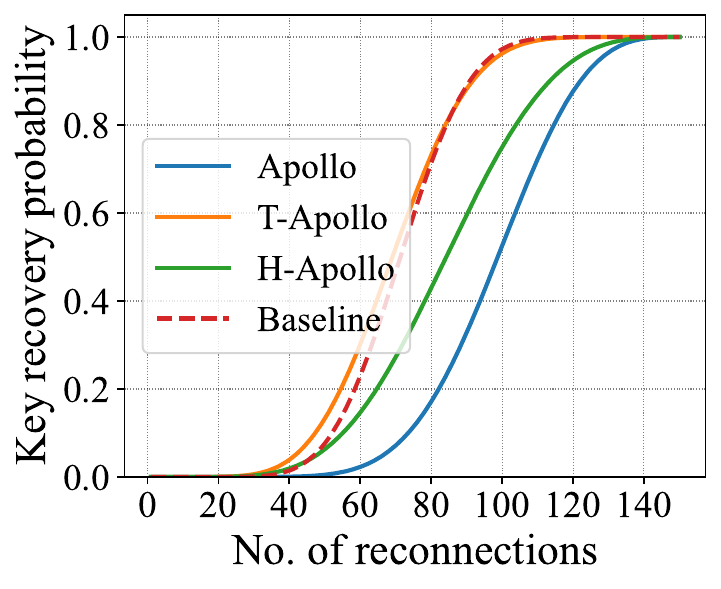}%
        \label{fig:app:eval:prob:variants}}
        \caption{Variation from single-layered and
        variants.}
        \label{fig:app:eval:prob}
\end{figure}

~\cref{fig:app:eval:prob:variation} depicts the variation of probability
from the single-layered approach (dashed curves).
The variation increases with the increase in address book size.
To reduce this variation,
we analyzed two variants of Apollo.
Thresholded Apollo or T-Apollo
uses Shamir's secret shares for subsecrets,
instead of additive shares.
Hinted Apollo or H-Apollo,
stores \emph{hints} (trustees in our case)
inside the backed up packets.
\cref{fig:app:eval:prob:variants} shows how
the variants bring the distribution closer to
the single-layered approach (baseline).

\section{Computation Complexity}\label{app:computation-complexity}
Here, we provide the bounds for combination operator.
\noindent \textbf{Lower bound to ${n \choose i}$:}\\
Let $0 < m < i \leq n$, then a simple computation gives $\frac{n}{i} \leq \frac{n-m}{i-m}$, as $i \leq n$ implies the following equivalences:
\begin{align}
   \frac{n}{i} &\leq \frac{n-m}{i-m}, \label{eq:lower-bound-comb-1}
\end{align}
This gives us:
\begin{align}
    {n \choose i} 
    &<
    \left(\frac{n}{i}\right)^i
\end{align}

\noindent \textbf{Upper bound to ${n \choose i}$:}\\
We have:
\begin{align}
    {n \choose i} &= \frac{n!}{(n-i)!\ i!} \leq \frac{n^i}{i!}.
    \label{eq:upper-bound-comb-1}
\end{align}

The power series expansion of $e^k$ for real $k$ gives us the relation $e^i > \frac{i^i}{i!}$, since
\begin{align*}
    e^k = \sum_{i=0}^{\infty} \frac{k^i}{i!}
\end{align*}
for any $k$, in particular for $k=i$. This leads to the equivalences
\begin{align}
    e^i &> \frac{i^i}{i!} \notag \\
 \iff   \frac{n^i}{i!} &< 
    \left(\frac{ne}{i}\right)^i.
\end{align}
Combined with~\cref{eq:upper-bound-comb-1}, we have an upper bound, given by
\begin{align}
    {n \choose i} 
    &< \left(\frac{ne}{i}\right)^i. \label{eq:upper-bound-comp}
\end{align}

\section{Practical Considerations}
\subsection{Wallclock Computation Time}
\label{app:practical:wallclock}

\begin{figure}[htb]
\centering
    \includegraphics[width=0.25\textwidth]
    {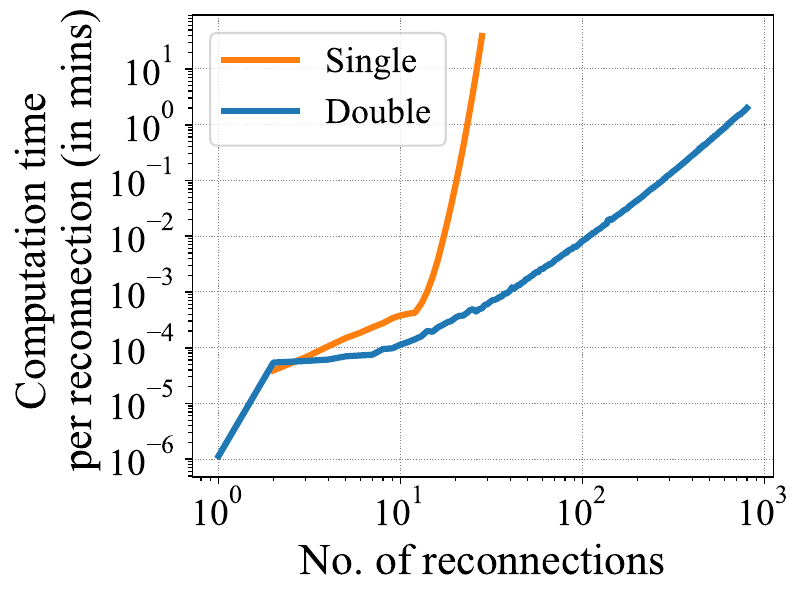}
    \caption{Wallclock time: time taken per reconnection.}\label{fig:wall:per}
\end{figure}

\cref{fig:wall:per}
depicts the wallclock time taken per reconnection.
The trends of these plots are essentially the
same as the plots with CPU time but the values are
different because of the parallelization of the implementation.

\section{Extensions of \name}
\label{app:practical:variants}
This section presents three extensions of \name:
Apollo Lite, Thresholded-\name and Hinted-\name
that make the recovery process easier for $\user$.
In \multiss, $\user$ generally has to contact
more than a threshold fraction of her trustees to reconstruct $\key$,
thus demanding more effort from her.
Apollo Lite reduces the threshold of each subsecret
by increasing the number of shares per subsecret.
This increases the number of relevant shares distributed
and brings the probability variation closer to
that of the single-layered approach.
On the other hand,
Thresholded-\name or \tname uses threshold in the subsecrets layer
to reduce the number of contacts approached.
Hinted-Apollo or \hname adds encrypted hints to the blobs
to make the approach of contacts less arbitrary.
Next, we describe the design of these variants and
their implications for security.

\subsection{Apollo Lite}
\label{app:practical:lite}
\begin{figure}[htbp]
    \centering
        \includegraphics[width=0.25\textwidth]
        {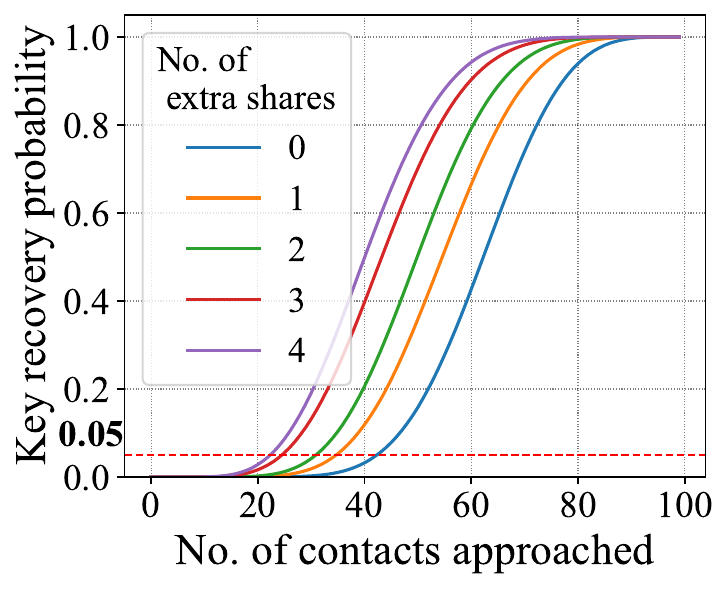}%
        \caption{Reduction in the no. of connections needed}
        \label{fig:eval:expected-prob}
\end{figure}

This extension is for shifting the expected probability
of recovery to that of
a single-layered approach.
For this purpose, the threshold of the subsecrets needs to be reduced and
in \name,
we can achieve this by increasing the number of shares per subsecret.
This approach would increase
the probability of key recovery at the threshold set
by the user, but it also introduces cases where the key can be
recovered by contacting less than threshold number of people.
As depicted by \cref{fig:eval:expected-prob}, the amount of
effort from user reduces as the number of extra shares increases.

\subsection{T-Apollo}
\label{app:thresholded-mlss}
Thresholded-Apollo or \tname uses a modified version of MLSS,
which we refer to as thresholded-MLSS or \tmultiss.
The design of \tmultiss is similar to that of \multiss,
except that \tmultiss uses Shamir's secret shares in the subsecrets layer.
Using a threshold for subsecrets is not equivalent to using a
smaller number of subsecrets
since the former provides higher share availability
by introducing redundancy in the shares distributed.
Thus, such a design serves two main purposes in key recovery.
First, it increases the probability of key recovery after
contacting $\lfloor {\threshold} \numtrusteesset \rfloor$ people,
thus mitigating the non-recovery issue
in MLSS (\cref{fig:multi-layered-fail}).
Second, \tmultiss enables key recovery even when
some contacts in the address book are unresponsive.
The trade-off is that \tmultiss increases the probability of
adversarial recovery via coercion, relative to the additive
secret sharing used for subsecrets in \multiss.
We now describe the design of \tmultiss.

\paragraph{Share distribution}
The share generation algorithm of \tmultiss is similar to
that of \multiss~(\cref{sec:apollo:mlss}),
differing only in the way subsecrets are generated.
In \tmultiss, subsecrets are Shamir's secret shares of $\key$
under a percentage threshold,
which can either be a default system value or be set based on
$\user$'s expected threshold.
Each subsecret is thus a coordinate pair, whose x-coordinate is
picked at random and whose y-coordinate is the corresponding
evaluation of the secret-sharing polynomial.
For each subsecret, Shamir's secret shares are then generated---%
each a coordinate pair, evaluated from a fresh polynomial whose
constant term is the subsecret's y-coordinate---%
and these shares are distributed among the \trustees.
Random shares keep the same form as in \multiss:
a coordinate pair with both coordinates chosen at random,
so that the pair does not lie on any of the subsecrets' polynomials.
Furthermore, the method for assigning key shares and random shares
among trustees to maintain indistinguishability
remains the same as in \multiss~(\cref{sec:apollo:mlss}).

\paragraph{Packet structure}
\begin{figure}[!t]
    \centering
        \includegraphics[width=0.48\textwidth]
        {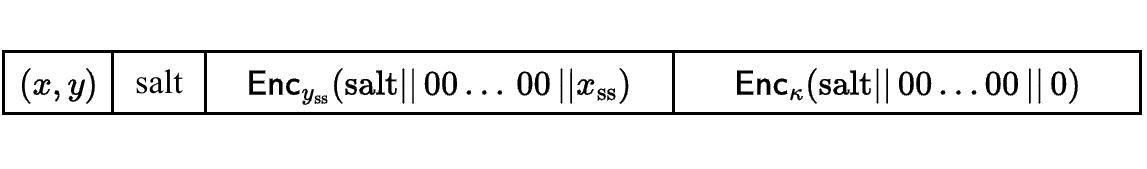}%
        \caption{Structure of a blob in Thresholded-MLSS}
        \label{fig:blob-tmlss}
\end{figure}
In \tmultiss, each subsecret is a Shamir share of $\key$ and thus
consists of an x-coordinate and a y-coordinate.
The share distribution backs up Shamir's secret shares of the
subsecret's y-coordinate, but carries no information about its
x-coordinate.
A blob must therefore store the subsecret's x-coordinate,
without breaking the indistinguishability of blobs.
Storing this x-coordinate in plaintext would leak structure:
two blobs holding shares of the same subsecret would carry the
same x-coordinate, letting an adversary link them to the same
subsecret.
Moreover, since blobs in \tmultiss must store recovery information
rather than merely indicate successful recovery as in \multiss,
storing a salted hash (as \multiss does) no longer suffices.

Instead, \tmultiss encrypts the \checksum together with the
information needed for recovery.
Each encryption stores the x-coordinate corresponding to the
recovered y-coordinate, and at the same time serves as an
indicator of successful recovery.
Because a deterministic cipher would let an adversary tell whether
two blobs hold shares of the same subsecret,
each blob includes a salt that randomizes its encryptions.
A blob signals successful recovery of a subsecret by padding the
encrypted value with the salt and a run of zeros; the recovery of
$\key$ itself is indicated by treating its x-coordinate as zero.
Concretely, a blob holding a key share stores, for each such share,
the encryption of the subsecret's x-coordinate---keyed on the
subsecret's y-coordinate and prefixed with the salt and padding---%
along with a corresponding encryption for $\key$.
\cref{fig:blob-tmlss} depicts this structure.
For random shares, a blob instead stores two random strings of the
same length as these encryptions.
As in \multiss, a trustee's packet is the set of its blobs.
Assuming the encryption scheme is semantically secure,
every blob is indistinguishable from every other,
thus making the \trustees indistinguishable from the \nontrustees.

\paragraph{Key Reconstruction}
\begin{algorithm}[!t]
    \caption{$\mathsf{Reconstruct}_{\tmultiss}$}\label{alg:reconstruction-tmlss}
    \begin{algorithmic}[1]
    \State \textbf{Input}: $P_{\obt} = \{\packet{i}\}_{i=1}^{k}, \alpha$
    \State \textbf{Output}: $\key_{\rec} \in \{\key, \nullval\}$
    \State $K = \{\share{i}\} \gets \mathsf{ExtractShares}(P_{\obt})$
    \State $E \gets \mathsf{GetEncryptionDict}(P_{\obt})$
    \State $R \gets \mathsf{GetNonceDict}(P_{\obt})$
    \State $\key_{\rec} \gets \nullval, \recovered \gets \mathbf{false}, 
    S \gets [ \ ]$
    \State $C = \{\{\share{i}\}_{i=1}^{\alpha}\} \gets 
    \mathsf{GetCombinations}(K, \alpha)$
    \For{$c \in C$}
        \State $y_{\interp} \gets \mathsf{Interpolate}(c)$
        \State $E_{\relev} \gets E[c[0]], \salt \gets R[c[0]], 
        \key_{\obt} \gets \nullval$
        \For{$e \in E_{\relev}$}
            \State $p = \mathsf{Dec}_{\mathrm{AES}}(y_{\interp}, e)$
            \If {$\mathsf{StartsWith}(p, \salt\, \vert\vert\, 00 \dots 00)$}
                \State $x_{\interp} = 
                \mathsf{GetXCoordinate}(\mathrm{plaintext})$
                \State $\ssmath{\interp} \gets (x_{\interp}, y_{\interp})$
                \If{$\ssmath{\interp} \notin S$}
                    \State $\mathsf{append}(S, \ssmath{\rec})$
                    \State $\key_{\obt} =  \mathsf{Interpolate}(S)$
                \EndIf
            \EndIf
        \EndFor
        \If{$\key_{\obt} \neq \nullval \text{ and } 
        \mathsf{len}(S) \neq 1$}
        \For{$e \in E_{\relev}$}
        \State $p = \mathsf{Dec}_{\mathrm{AES}}(\key_{\obt}, e)$
            \If {$\mathsf{StartsWith}(p, \salt\, \vert\vert\, 00 \dots 00)$}
                \State $\key_{\rec} \gets \key_{\obt},
                \recovered \gets \mathbf{true}$
                \State \textbf{break}
            \EndIf
        \EndFor
        \EndIf
        \If{$\recovered = \mathbf{true}$}
            \State \textbf{break}
        \EndIf
    \EndFor
    \State \textbf{return} $\key_{\rec}$
    \end{algorithmic}
\end{algorithm}

The key reconstruction algorithm of \tmultiss is outlined
in~\cref{alg:reconstruction-tmlss}.
As in \multiss,
\tmultiss does not expect the user to remember her set of trustees
or the thresholds used.
Key reconstruction in \tmultiss resembles that in \multiss,
differing only in the \checksum verification
and in how $\key$ is reconstructed from subsecrets.
Specifically, instead of matching salted hashes of the subsecrets,
$\user$ decrypts the blobs' encryptions to detect correct recovery.
When the decryption reveals the blob's salt and the padded zeros,
the algorithm recognizes successful recovery of the subsecret
and reads off the corresponding x-coordinate.
Moreover, since subsecrets are Shamir's secret shares of $\key$,
\tmultiss reconstructs $\key$ from them by
polynomial interpolation rather than addition.

\paragraph{Security Implications}
With \tname, $\user$ reconstructs fewer subsecrets for the same
number of contacts holding shares as in MLSS,
because each subsecret now requires only a threshold fraction of its
shares rather than all of them.
Consequently, the minimum number of trustees that must be contacted
also decreases.
Thus, \tmultiss reduces the effort required from $\user$
during reconstruction,
at the cost of a higher probability of coercive recovery
compared to \multiss.
\cref{fig:eval:variants} compares \tmultiss against \multiss.

We now describe the design of \hname.

\subsection{H-Apollo}
Hinted-Apollo or \hname modifies
how $\user$ locates shares in her address book during recovery.
Specifically,
\hname stores \emph{hints} in the blobs held by \trustees,
each naming a trustee who holds shares of another subsecret.
The hints are encrypted under the subsecrets as keys,
so they guide $\user$ as she proceeds with the recovery.
Thus, \hname reduces the effort $\user$ must expend
to obtain shares from her address book,
and thereby increases the
probability of key recovery after contacting
$\lfloor {\threshold} \numtrusteesset \rfloor$ \trustees.
Conversely, the hints could aid $\adversary$ during coercive
recovery, yielding weaker security than \name.
\hname uses \hmultiss, a modified version of \multiss,
which we describe next.

\paragraph{Share distribution}
\begin{figure}[!t]
    \centering
        \includegraphics[width=0.48\textwidth]
        {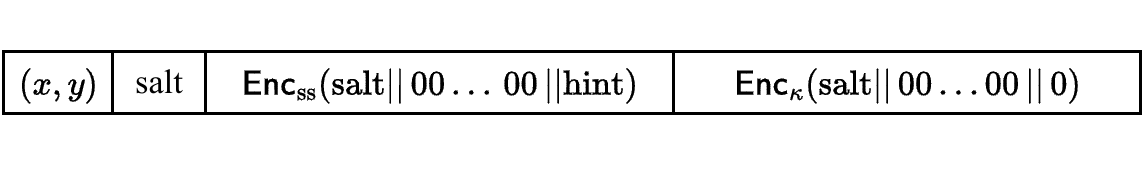}%
        \caption{Structure of a blob in Hinted-MLSS}
        \label{fig:blob-hmlss}
\end{figure}
The share generation in \hmultiss is similar to that in \multiss.
The subsecrets are additive shares of $\key$,
and the key shares are the Shamir's secret shares of the subsecrets.

\paragraph{Packet structure}
In \hname, hints are revealed to $\user$ as she reconstructs more subsecrets.
Each hint is therefore encrypted and stored in the blobs held by \trustees,
using the subsecrets as encryption keys.
The packet structure in \hmultiss is the same as in \tmultiss,
differing only in what the encryption contains:
instead of an x-coordinate, each blob stores the name of a trustee.
As in \tmultiss, the encryption corresponding to $\key$
stores a string of zeros in place of a hint.
\cref{fig:blob-hmlss} depicts a blob's structure in \hname.

A second parameter governing \hname's security
is the total number of hints stored across all blobs.
Backing up the names of all \trustees naively across every blob
would raise the chances of coercive recovery,
since recovering one subsecret would then point the adversary
to the remaining \trustees.
\hname therefore caps the number of \trustees that the blobs reveal
as a whole.
If only one trustee is to be hinted,
every blob stores the encryption of that single trustee's name.
When more hints are needed,
\hname first fixes how many trustees may be revealed,
samples that many names uniformly at random from the address book,
and then assigns each key-share blob a hint drawn uniformly at random
from this sampled set.
The cap on revealed \trustees is chosen based on $\user$'s
expected threshold.
For random shares, the blobs store no hints,
holding instead two random strings of the same length as the
encryptions in key-share blobs.

\paragraph{Key Reconstruction}
The key reconstruction algorithm in \hmultiss is
identical to that in \tmultiss (\cref{alg:reconstruction-tmlss}).
The difference lies in how $\user$ approaches people in her
address book and in how $\key$ is reconstructed from subsecrets.
Rather than approaching contacts in an arbitrary order,
$\user$ follows the hints in the blobs she recovers.
Since she reaches more \trustees as she recovers subsecrets,
the overall computation time also decreases relative to \multiss.
Moreover, since the subsecrets are additive shares of $\key$,
$\key$ is reconstructed by adding the subsecrets.

\paragraph{Security Implications}
In \hname, the hints let $\user$ contact fewer people.
While this makes recovery easier for $\user$,
\hname also leaks more information to $\adversary$ than \name does.

\subsection{Comparison with Apollo}
\begin{figure}[!t]
    \centering
        {\includegraphics[width=0.25\textwidth]
        {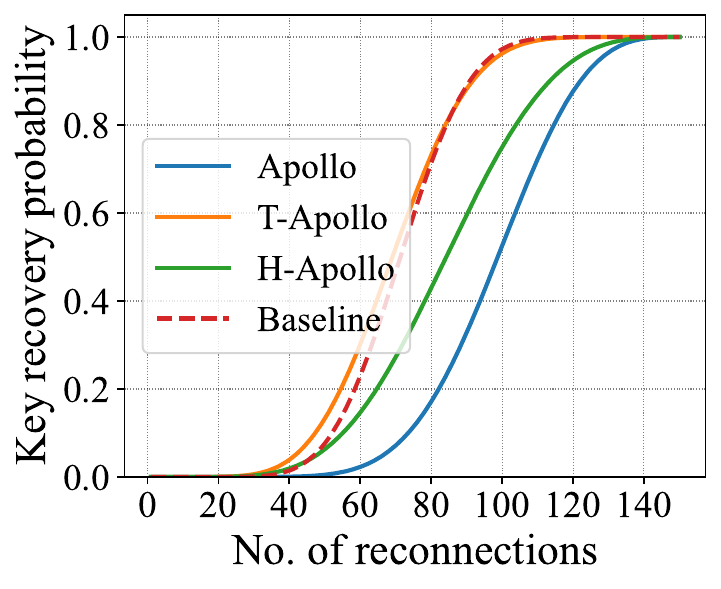}%
        \label{fig:eval:variants:probability}}
        \caption{Comparison with variants (probability of key recovery): TMLSS with $80\%$ threshold
        in the subsecrets and HMLSS with $5$ hints}
        \label{fig:eval:variants}
\end{figure}

In~\cref{fig:eval:variants}, we compare the probability of key recovery using Apollo and the variants described above.

\end{document}